BERNARDO SORJ

# INFORMATION SOCIETIES AND DIGITAL DIVIDES

an introduction

Polimetrica®

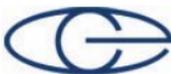

**the edelstein center for social research**
**www.edelsteincenter.org**

PUBLISHING STUDIES

directed by Giandomenico Sica

VOLUME 4

**BERNARDO SORJ** [1]

# INFORMATION SOCIETIES AND DIGITAL DIVIDES

## an introduction

**Note for the Reader**

In our view, doing research means building new knowledge, setting new questions, trying to find new answers, assembling and dismantling frames of interpretation of reality.

**Do you want to participate actively in our research activities?**

**Submit new questions!**

Send an email to the address **questions@polimetrica.org** and include in the message your list of questions related to the subject of this book.

Your questions can be published in the next edition of the book, together with the author's answers.

**Please do it.**

**This operation only takes you a few minutes but it is very important for us, in order to develop the contents of this research.**

Thank you very much for your help and cooperation!

We're open to discuss further collaborations and proposals.
If you have any idea, please contact us at the following address:

*Editorial office*
*POLIMETRICA*
*Corso Milano 26*
*20052 Monza MI Italy*
*Phone: ++39.039.2301829*
*E-mail: info@polimetrica.org*

**We are looking forward to getting in touch with you.**

# LIST OF QUESTIONS

# INTRODUCTION

The topic of this book, the digital divide, refers to the unequal distribution of resources associated with information and communication technologies, between countries and within societies. The case of access to information technologies, provides an excellent opportunity to study the way that one technology-based product, in this case the Internet, can favor both economic and social development, greater freedom and circulation of information and social participation while it also possesses the capacity to deepen social inequality and create new forms of power concentration.

To approach the challenge of analyzing the digital divide, we must avoid simple, easily formulated views with strong media and/or political appeal, which overlook the richness, diversity, and complexity of social life. For some authors and international institutions, new technologies may allow less developed countries and poorer sectors of the population to substitute advanced technologies for investment in education, producing economic and social leapfrog. Others argue that new technologies will widen the gap between the rich and the poor both internationally and within societies hence the digital divide is a secondary problem, and that new technologies are luxuries of a consumer's society.

These visions reveal real but only partial tendencies. While current data shows that new information technologies tend in general to increase social inequality, giving new advantages to the more educated, there are indications that they could be equally powerful in helping the least privileged sectors of the population.

The impact of each technology depends on the way it is creatively appropriated by different social groups, in specific economic and political conditions. Dismissing new technologies represents a narrow and elitist perspective of the consumer world. Though we accept that new technologies are not a panacea for the problems of inequality, their universalization is today among the fundamental conditions of integration within society. Therefore the social consequences of new technologies are neither linear nor predictable and are capable of generating new forms of stratification and social fragmentation. To understand the digital divide it is necessary to place it within the broader dynamic of each society and the international system, as an element in the set of goods and services that determine social inequality. The shape of the information society will depend on the way new technologies are creatively appropriated by companies, non-governmental organizations, social movements and public policies. Therefore instead of a single model of *information society* we prefer the notion of *information societies*, as each society absorbs in different ways the new technologies.

Communication and information technologies (C&IT) include an array of products (radio, TV, cable TV, fixed and mobile phones, computers, Internet access), and all of them are part of the information society and each one of them creates its own digital divide. In this study, we will focus on the digital divide related to the social, economic and cultural consequences of the unequal distribution of access to the Internet. We will need focus therefore on the specific consequences of the new technologies on the economy. The text does not look specifically at radio, TV, fixed and mobile telephone access.[2] Although these technologies belong to the same set of information communication technologies, even sharing the same infrastructure, from a sociological perspective, these products have quite different qualities and each of them produce a specific digital divide. The "old" communication technologies are part of the family of 'illiterate friendly' products – that is, products that can be used by individuals who have very low reading and writing skills – while computers and Internet demand basic educational skills. However, in the future, the convergence of technologies will increase the need to process written information even to manipulate a cell phone creating an internal digital divide among users according to their literacy.

Technology embedded goods are massively consumed and therefore we need to integrate them within a larger view of the meaning of consumption in contemporary societies. In the first chapter I argue that although there are consumer objects with mainly symbolic dimensions, from which people establish social status, and although product choice can be influenced by publicity and personal taste, most of the technology embedded consumption products are prerequisites of civilized life, of access to a better quality of life, to jobs, and to active participation in society. A large part of the bibliography concerning the consuming society is focused on a relatively small number of fashionable products saying little about the majority of products consumed in contemporary society. This includes products that are not oriented toward a specific social class and that are not consumed because, or not mainly, of the influence of publicity or pure individual choice. New consumer goods are scientifically embedded technological artifacts that have profoundly transformed so-ciety and the access to these goods is a prerequisite for full participation in social life.

Inequalities are both the product of different personal income but they are also related to the distribution policies of the state. In the second chapter we present a general view of the multiple aspects of social inequality and their relationship to private and collective consumption goods. The various kinds of social inequality – personal and collective – are not independent; they are interrelated and reinforce or neutralize one another. The digital divide has added another dimension to the diversity of existing inequalities in society: that of unequal access to the set of new goods and services associated with new information and communication technology.

Beginning with a summary description of the principal dimensions of the information society, chapter three will show that an emphasis on new processes does not allow us to overlook the continuity that exists in social organization. After presenting the general impact of Internet on society we argue that in spite of the expansion of networks, pyramid power shaped structures continue to control vast resources of power. The challenge for democratic change in the contemporary world is the integration between networks and pyramids, between states and non-governmental organizations, and between national and international organizations.

In the fourth chapter we introduce the problem of the digital divide and how it is related to the overall problem of social inequality, economic development, and the fight against poverty. The digital divide is played out on many levels, in each case with specific impacts on social inequality. The existence of physical infrastructure, access to individual connections, digital literacy, education, and contents developed specifically to reach the needs of the poorest sectors of the population all have an impact on inequality.

At the conclusion we analyze the principal problems that must be addressed by public policy to fight the digital divide, particularly in developing countries, to establish efficient uses of resources. Public policy and civil society initiatives should confront the diverse forms of social inequality as an interdependent set of phenomenon and address them with simultaneous and coordinated actions.

Given the continuously changing situation, we prefer instead of including statistics and graphics in the text to refer the reader at the end of the book to the sites were he/she can find updated statistics. Finally I would like to thank Joel Edelstein for his comments on a first draft and Giandomenico Sica for his invitation to write this book.

# 1. CONSUMPTION

## *1.1 How should we analyze consumption?*

Social science lacks comprehensive theories regarding the relationship between consumption and society. Economics, with its nearly exclusive focus on efficiency and rationality has reduced consumption to a simple question of personal choice and utility. Sociological theories, such as those mentioned earlier, were developed to confront this perspective showing that consumer options are constructed socially and that individuals make choices based on aesthetic standards and life styles. However, most of these theories have over-emphasized the cultural dimensions, leaving aside the material reality of consumer goods.

In recent decades theories regarding the consumption society have focused on the analysis of consumption objects as a symbolic system, a type of communication system in which consumption objects main role is to be used as symbols of social status. Thus the use value of a product is viewed not in its material utility, but rather by its capacity to indicate social distinctions. The differentiation of products are related to the differentiation between social groups, in particular middle and upper classes make a constant effort to mark their social status through objects and an aesthetic to which only they can afford.

The main supposition of many of those critical of the consumption society is based on the idea that there are "real" as against "created" (or artificial) needs. In contemporary society in particular, most goods would be based on artificial needs, reflecting the demand created by publicity and of a society dominated by exhibitionism and ostentation. However anthropological studies have criticized the notion that there is a direct link between what we consume and intrinsic or natural need. Culture always permeates tastes and defines which products are appropriate for consumption. For example, there is nothing natural about the readiness in some cultures to eat beef while others prefer dog meat.[3] Anthropology also shows that consumption objects are not only instruments of social distinction for dominant groups, since they are used equally to mark a variety of identities – of minorities, underprivileged, age or protest groups –.[4]

## 1.2 What are consumption goods?

**KEYWORDS:**
CONSUMPTION, TECHNOLOGY EMBEDDED PRODUCTS

Though ostentation has always characterized, and will continue to characterize particularly the lives of the dominant classes, in modern society most goods are not consumed due to publicity but because the large majority of consumption products are pre-conditions of access to health, education, work, and sociability. The majority of consumption products are used because, within the context of contemporary society, they are useful. They represent technologies that allow improvements to the quality of life and social integration. After they have reached a certain level of dissemination in society, not having them means social ostracism. For example, not having a phone, or more recently an email address, can lead to social exclusion.

Theories that connect social stratification with consumption and describe its symbolic aspects tend to forget that this dimension has, in general, only secondary importance in the world of consumption. This also includes economics with its focus on production. These theories ignore the central char-acteristic of consumption goods in society: **from the beginning of mankind knowledge was embedded as much in the objects of consumption (food, housing, medicines) as in the means of production. And changes in the ways of consumption have major consequences on social relations and life expectancy, affecting the whole fabric of society including the economy**. In contemporary societies the majority of consumption products are embedded expressions of scientific and technological knowledge, and did become prerequisites for social integration in everyday life, as much in terms of quality of life as in terms of chances for participation in society in general and in the labor market in particular.

Consumption as a mechanism of social distinction in a mass society is important at the margins of the productive sector. Much advertising focuses primarily on brand fidelity. In fact many consumption products like home water, electricity, radio, TV and telephone lines, which were also once objects of conspicuous consumption, are now considered commodities or even public goods. The critiques of the McDonaldization of the world and the alienating role of trademarks epitomized by Nike refer only to one side, which is generally secondary, of the world of consumption. For the poorest populations of the planet, globalization is not so much the expectation of eating at McDonald's or wearing Nike, it is access to food, water, electricity, appliances, radio, television, telephone, Internet, antibiotics, books, cinema, CD players, cars, travel, and all those products and services to which persons opposed to globalization would not deprive themselves. These products bring material quality of life to such a level that the poorest person in France today enjoys a better material quality of life than richest French person did 200 years ago. Be it to treat a toothache or an infection, to access information, to deliver children, to lie down in a comfortable bed, or to have a good heating or cooling system to face winter or summer, con-sumption goods are disseminated because they facilitate life, not through publicity brainwashing.

Capitalist consumption civilization's principal challenges are not so much the addiction to given brands but in their impact on environmental and the relationship between consumer goods, intellectual property, privacy and ethics. Such is the case, for example, with the problems surrounding genetic engineering which raises questions about the control of life or, as we will see, information technology with its potential to destroy privacy.

## 1.3 What are the social consequences of consumption?

**KEYWORDS:**
CAPITALISM, MARX, PARADOXICAL CONSEQUENCES

Consumption products, particularly within capitalist society, but also throughout all human history, incorporate and condense as much technology and knowledge as the instruments used in the production process. Social relations are transformed through consumption as much as through relations of production.

The difficulties that social science encounters as it attempts to analyze the role of consumption in capitalist civilization originate with a false starting point that radically separates the processes of production and consumption. Treating both spheres as autonomous realities results in a search for separate explanations for each one. But consumption is one of the constitutive elements of the productive system, not only because it allows the flow of production but also because the majority of products consumed transform the context of production, workers life style, productivity and social relations.

Even Marx, who tried to relate production and consumption, made a major analytical error by reducing work to a commodity. He didn't grasp that labor is a very particular commodity, one that not only is capable of fighting for its market value, but which is transformed through the process of consuming other commodities. In other words, consumer goods have qualities that affect the way society is organized. To the degree to which they increase longevity, facilitate locomotion, permit communication between workers and their private lives, and reduce time spent on housework, they directly impact the productive system, transform society, and become essential to social life.

The social consequences of consumption cannot be reduced to the specific utility that a product brings its users, given that there are often other consequential or paradoxical impacts when they are disseminated throughout society. In many cases, the dynamics of the social product cycle play a role in determining the product utility. The diffusion of a product is in some cases a condition of its practical use. This is due either to the availability of necessary infrastructure, as is the case with paved roads and highways whose development depends on a minimum number of cars in circulation; or the existence of other users to permit interaction, as with case of the telephone, which requires a minimum number of telephone owners to become a useful tool of communication.

The paradoxical effects are the product of unexpected dysfunctions produced by the dissemination of a given product or service. For example, the car facilitates transport of an individual from one place to another, but today in many cities, an excess of cars makes the bicycle a faster mode of transportation not to mention its non-polluting nature. New medicines can save lives but their effects on the human gene pool could be disastrous. New seeds can increase production but can also have irreversible negative impact on biodiversity. Or, as we will show, with the case of the information society, databases containing patient medical histories or credit card purchase records can save lives or reduce the risks involved with carrying money, but they gen-erate information on people's private lives that could lead to control over privacy and new types of discrimination based on genetics or illness expectancies (at work or insurance).

## *1.4 What are the "social cycle of the product cycle" and the "international social cycle of the product"?*

**KEYWORDS:**

SOCIAL CYCLE OF THE PRODUCT, INTERNATIONAL SOCIAL CYCLE OF THE PRODUCT, STRATIFICATION, INTERNATIONAL INEQUALITIES

The social dynamics of new technologically based consumption goods goes through a **social product cycle** and an **international social product cycle**. The **social product cycle** is based on the economic product cycle, in which initially a product is introduced in small quantities with high price and later it is mass-produced permitting access of a large part of the population.

A new product initially reaches only those at the highest income levels and later, with mass production and price reduction, is disseminated throughout all sectors of the population. Thus, the dynamics of technological innovation reinforce social inequality in the initial stages, when a product reaches only the highest income sectors of the population, later to play an equalizing role through mass production.

The **international product cycle refers** to the dissemination of new products, mostly created in advanced countries, on a global scale. In developing countries many products reach the highest classes and later the middle classes, but in many cases it takes a long time to reach the lowest classes and in some cases never do. When the product cycle is incomplete, excluding certain sectors of the population from new technological innovation, it consolidates and/or creates new forms of social inequality.

## 2. SOCIAL INEQUALITY

### 2.1 What are the forms of social inequality?

**KEYWORDS:**
INEQUALITY, INDIVIDUAL INCOME, INDIVIDUAL AND COLLECTIVE PUBLIC GOODS, GLOBAL SOCIAL GOODS

Social inequality supposes differential access to social wealth within a social system. Analyses of social inequality traditionally distribute the population of a country as though it were a continuum of individuals, a straight line from those who have a lot to those who have very little. Studies on social inequality analyze the distance between the poorest and richest sectors of the population taking as an indicator the income of individuals or families. Individual income is without question an important criterion in social inequality but it represents only one dimension of the problem. The unequal distribution of public goods and services is equally important, and in some cases this unequal distribution is even more decisive in defining life chances. Not being able to count on police protection or access to electricity, water, sewage and telephone networks, medical services, or schools, has dramatic consequences on the quality of life. These goods are delivered, directly or indirectly, through the public sector, state concessions or publicly regulated services.

Goods and services can be divided into two major categories: individual and collective consumption goods. Individual consumption goods are those that are selected on the basis of individual personal option. Goods and services for collective consumption are those that, in a given historical period, are considered fundamental conditions for citizenship and therefore require public intervention to ensure universal access.

Public action concerning collective goods can cause them to lose the quality of being mercantile goods. In other cases
they can be produced and/or distributed through the markets,
under public control. In all of these cases the state intervenes,
be it to through orienting investments or subsidies and price
controls, to insure universal access – independent of individ-
ual income –. Though economists have tried to identify traits
or qualities that can be associated with public utility or social
goods, there are no criteria to distinguish public and private
goods in their pure states. The definition of the public or so-
cial character of goods will depend on the values of each so-
ciety in a given historical moment. In democratic societies,
public debate determines which products and services should
be universally accessible.

In the past few years, a new type of collective consump-
tion goods has begun to be discussed **global social goods**.
These are goods that cannot be delimited by national borders
or whose absence in one country can affect the quality of life
in other countries (such as the protection of the environment,
control of epidemics, but also free international circulation of
ideas and information). The concept of global social goods
and the unequal access to them has yet to be adopted and
elaborated by public opinion, but they have an enormous
practical and political potential as they demand a broad dis-
cussion on global governance mechanisms that can ensure ef-
fective creation of an international space for public goods

*2.2 What are the types of collective goods?*

**KEYWORDS:**
BASIC SERVICES, COMMON SERVICES, INDIVISIBLE GOODS, NETWORKS,
NATURAL RESOURCES

There are four types of collective consumption goods in con-
temporary national societies. The first group is made up of public services connected to basic government institutions. Access to these services is completely separate from payment. They are state run goods and services like police, the justice system, and the services associated with the functioning of the legal system. Their nature requires a complete separation between public employees and citizen's income or ability to pay, to ensure egalitarian and universal treatment. The financing of these services is carried out indirectly through the tax system.

A second group is made up of common services such as public lighting, cleaning services, road maintenance, parks and gardens, fire department, and environment. Generally they are the responsibility of local authorities. They are financed through taxes, usually taxes related to housing, and can be carried out by public or private companies under state concession contracts and regulation.

A third group is made up of collective goods and services that are not by nature indivisible. This is made up of goods and services that are regarded as basic conditions of citizenship such as health, education (at least basic education), and social security and, to a lesser extent, housing. They are financed by direct taxation and/or employers and employees contributions. These goods and services can be offered by public or private institutions as well as by non-profit organizations. When they are offered directly by the state, these services generally coexist with private services of the same type oriented toward individual consumers such as private health plans and private education.

Finally, the fourth category of collective consumption goods is made up of goods and services considered essential or of public interest that are connected to **networks and/or natural resources** that control a given space (be it under the earth in urban areas, or in the form of waves that travel through air) and give their owners a position of monopoly or oligopoly. They include water, electricity, sewage, radio, television, transportation, and telephone. These collective con-sumption goods can be offered either by public or private companies with use rights delegated by the state. In either case, the state takes responsibility for ensuring the quality and pricing of these services and that the network's controllers provide universal access. Though these services are generally paid by individual consumers (or in some cases by taxes), effective access requires that the network reaches every home and that the prices are reasonable even for low-income groups. While in the case of the previous types of collective consumption goods, it is expected that the different capacity of each citizen to contribute financially to the goods and services is carried out in a progressive way – e.g. a taxation structure where fees are inversely related to income –, in the case of networked goods and services, compensatory pricing systems usually take the form of subsidies in which certain consumers (for example, those in high income areas or businesses) pay a higher price thereby permitting other consumers to pay a reduced price for the same services.

These state regulated goods and services impact positively on social inequality by partially or totally dissociating access to goods and services from personal means. To these public services should be added other redistribution policies, which include unemployment insurance, social services, disability insurance, minimum income policies, food distribution programs, and subsidized housing.

## *2.3 How can social inequality be confronted?*

**KEYWORDS:**
MARKET, STATE, INDIVIDUAL INITIATIVE, SOLIDARITY, EQUALITY, CITIZENSHIP

The fight against social inequality therefore takes place on two fronts: at the level of the market and at the level of state as provider of social goods, regulating the private sector. Contemporary capitalism therefore is the product of the com-bined and contradictory actions of two structuring principles: on the one hand liberty and individual initiative – consolidated in the institutions of private property, of freedom of contracts and of markets as organizer of the system of production and of exchange–; and on the other hand the values of solidarity and equality – expressed through the idea of citizenship and of the nation as a community that should ensure for its inhabitants some minimum living conditions and chances for participation in society–.

While the first principle indicates that each person should acquire goods and services in the market according to his or her assets and personal options, the second requires some private intervention in the system of distribution to assure that all citizens have access to a minimum set of goods and services that are considered basic at the particular point in history. In practice these two contradictory principles coexist in all capitalist societies and the results of the conflict and synthesis of these principles defines the specific profile of each national society. It is important to note that they are different principles, associated with values that coexist simultaneously in modernity and that there is no scientific formula that can resolve the problem of how to combine them. Each solution will depend on political struggle and on working out solutions and creative arguments with the capacity to convince the majority of society. The coexistence of these two organizing principles of capitalism constantly generates new models that define the pattern of state regulation and public intervention in the production, distribution, and consumption of goods and services.

Each new product that society considers a basic condition of social integration and quality of life becomes a field of struggle regarding the way its dissemination should be assured to all citizens, through public or private means, or a mixture of both. The case of information and communication technologies is a typical example of this type of product.

## *2.4 How does consumption relate to inequality?*

**KEYWORDS:**
NEW PRODUCTS, CIVILIZING THRESHOLD, COLLECTIVE AND INDIVIDUAL
GOODS, SOCIAL POLICIES

The central question for contemporary society is that the permanent introduction of new consumption products that impacts quality of life involves changes in the civilizing threshold, i.e. , which goods are considered essential for a decent life in society. In this sense, each technological innovation that introduces new consumption products changes the perception of what it means to be socially included or excluded (based on access or lack thereof), thereby changing the universe of goods and services that require some type of state intervention. In other words, poverty, and the fight against it are dynamic and require constant efforts to re-adapt social policies.

A classification of individual consumption goods is outside the limits of this work but it is fundamental to emphasize its deep relationship with collective consumption goods. The different individual and collective goods and services cannot be dissociated from one another. Personal health, for example, is enormously affected by the problems caused by lack of treated water and sewage, the principal causes of child illness in poor neighborhoods in developing countries. The large majority of individual consumption goods depend on prior access to collective consumption goods. To use home appliances, access to electricity is needed. Telephone use requires telephone access to telephone networks. Home water and sewage services require access to urban infrastructure.

The relationship between individual income and access to goods and services can generate both virtuous and vicious circles. For example people who live in neighborhoods dominated by drug trafficking find difficulty in getting jobs. Simi-larly, low-income levels force some families to take their children out of schools so that they can make an early entrance into the workforce. At the same time low incomes are often the result of low education levels that limit the possibilities for obtaining higher wage jobs. On the other hand, for instance, students from highly educated and well off families have more chances of access to public universities.

Combating the many dimensions of inequality requires a complex view of social policies. The challenge is to identify those areas where positive consequences can be greatest both in the short and medium term. Some actions, such as investment in education may be decisive to improve personal income but they take years to mature while other actions such as extension of water and sewage networks have immediate consequences over quality of life but do not directly impact on income levels. In practice, social policies are oriented both by technical logic and by pressures from diverse social groups in each city and state. In each location needs are different and the capacity of each social group, including the lower, middle, and upper classes, to put pressure on the state, determines the priorities of public investments.

## 2.5 What is the difference between the Internet and previous consumption goods?

**KEYWORDS:**
MASS CONSUMPTION, INTERNET, LITERACY, ACCESS BARRIERS

The new wave of information technology related to the Internet has characteristics that are new or more pronounced than in previous waves of mass consumption. First, information technology, in addition to being interactive (like the telephone), is **proactive**, meaning that it allows users to personally appropriate the contents of instruments of communication (for example by making their own web site). The uses and possibilities of information technology depend, however, on the intellectual training, in particular the education and profession of the user. While the use of appliances, telephones, radio, and television require almost no formal education, information technology not only requires literacy but its usefulness depends on each user's intellectual ability in selecting, analyzing, understanding, and evaluating available information. Although the Internet can influence a user's ability to analyze, we will see that these abilities are developed in large part outside the Internet. While for the user with limited analytical competence the Internet is an information tool, for the user with greater analytical capacity, it is a knowledge tool.

In addition, information technology presents specific barriers to initial access that are greater than those of previous electronic products. The majority of previous electronic products require minimal service expense after initial purchase. The only exception is the telephone with its monthly service charges. Internet requires either fixed monthly fees (for example wide-band service), or increases in telephone charges (for users of dial-up service, as flat-rate local residential service is not available in most countries). We will see that these fees make up one of the principal barriers for the diffusion of information technology among low income groups (and sometimes even the low middle class). In the case of the computer, their use requires other ongoing expenses such as printer cartridges, paper, periodic technical support, updates to programs, and the constant need to update equipment that makes these products reach obsolescence rapidly.

Finally, information technology products, due to their proactive character, are for individual and personal use. Older systems of communication such as radio, television, and fixed telephones, were easily shared between family members.

# 3. THE INFORMATION SOCIETIES

## *3.1 What is the information society?*

**KEYWORDS:**
INFORMATION SOCIETY, KNOWLEDGE SOCIETY, TECHNOLOGY CONSUMING
CAPITALIST SOCIETIES

The term "Information Society" is currently the most common
way to refer to the impact and social consequences of new in-
formation and communication technologies. While it is useful
as a concept that identifies a theme, it is not a theory or an ex-
planatory framework for the dynamics of societies in the con-
temporary world. In a strict sense, the term is also incorrect,
first because information is equally important in all societies,
and second because information on its own has no value, its
relevance depends on its insertion into a system of knowledge.
In this sense, another widely used term, "knowledge society",
is more appropriate, but once again the term overlooks the fact
that all societies are based on knowledge. In practice the con-
cept of a "knowledge society" refers to a particular type of
knowledge, scientific knowledge, through which technological
innovation, the principal vehicle for economic expansion in
the contemporary world, is possible. From a sociological point
of view, it is perhaps more appropriate to speak of **technology
-consuming capitalist societies**, that is, societies where com-
munication, quality of life, and economic, social and political
relations are mediated by technological artifacts (in the form of
products and services) that incorporate scientific knowledge
under capitalist based social relations.

Since the social processes associated with the "Informa-
tion Revolution" are in their initial phases, many analysts
confuse trends, extrapolations, and speculation with current
reality. Certain argumentative exaggerations play an impor-tant role in expanding our field of perception and sensibility to new phenomenon but it is important, especially with regards to the use of scarce resources in public policy, to focus as much on continuities as on discontinuities, on the new and the old, without carelessly extrapolating experiences from other contexts, remembering that the world is not California and that each land has its own realities.

## 3.2 What has changed with the Internet?

**KEYWORDS:**
KNOWLEDGE MANIPULATION, KNOWLEDGE, COMMUNICATION, VIRTUAL SPACE

The unilateral emphasis on the impact of the Internet can create a perception of a radical transformation dividing the new and old forms of social organization. But we cannot overlook the fact that the computer has been influencing society for several decades. Its influence was already discussed extensively in the seventies and eighties before the Internet. The Internet represents a new communication technology that adds, in a revolutionary way, to the long list of instruments of voice and image transmission such as telegraph, telephone, record-player, telex, radio, television, and fax that have changed communication in contemporary society.

Information technology, and its most widespread system, the Internet, are of enormous importance because they allow the convergence of two activities that are central to social life: the manipulation of knowledge and communication. Information technology allows the storage, organization, and processing of an enormous amount of information in a small space and at incredible speed. New communication technologies permit instantaneous voice, text, or image communica-tion on a worldwide scale, constantly increasing the availability of information while decreasing communication costs.

These combined technologies working through a set of protocols (TCP/IP is the most common on the Internet) allow communication between computers. The Internet is a network of computer networks, all communicating in real time, making information instantaneously available in any part of the planet. Thus, information and communication cease to be spatially localized and are transferred to "virtual space" (or cyberspace), allowing simultaneous contact between an infinite number of people using the memory of the computers participating in the network, independent of their physical location. While this technology is associated with computers as physical differentiated objects (desktop or laptop) convergence tendencies are transforming mobile phones or TVs into the new vehicle of access to Internet.

## 3.3 What is the social context of the information revolution?

**KEYWORDS:**

CHANGES IN CAPITALISM, SERVICE SECTOR, WORK FLEXIBILISATION, DEMATERIALIZATION, INDIVIDUALIZATION, GLOBALIZATION

The Internet appeared in a period when capitalism was undergoing a deep change in its productive and social system. It acted as a catalyser and accelerator, but the Internet wasn't the only cause for these transformations. By forgetting recent social and economic history, several authors have ended up with technological determinist interpretations. In order to get a historical perspective it is worth mentioning, albeit in a summary form, processes which although prior to and not correlated to the Internet, were a strong impetus to the new C&I technologies:

1) The transformation, in the last decades, of the service sector in the dynamic core of the productive system. The capacities for technological innovation associated with the application of knowledge became the principal source of aggregate value, productivity gains and dynamism of the economy. Information technology was not the initiator of the so-called information society or knowledge based society, but an accelerator or vector of a process that preceded it. The increasing importance of applied scientific knowledge as the principal source of innovation and value creation in a constantly changing world transforms learning into an ongoing process, driven by the necessity to update and adapt professional skills to the requirements of new technological transformations.

By putting a large part of human knowledge in virtual space, facilitating the interchange and expression of ideas and developing online services in real time, the Internet allows people to break the barriers that in the past have limited access to, and transmission of, information. But the Internet is not a substitute for human capital, which is the product of large, long-term, investments. Nor does it substitute the laboratories, research centers, and corporate resources under which scientific knowledge is produced and transformed into technology and finally into consumer products.

2) The increasing "flexibilisation" of the work process and the production arrangements. This trend is partly associated to the processes described above, in particular to the greater value assigned to knowledge demanding greater autonomy and creativity, and overall changes in the socio-political system with  the relative decline of trade unions, welfare benefits and labor rights. The Internet, in some cases has been a tool for the advancement of new models of management and the flexibilization and decentralization of production and work.

3) The tendency known as dematerialization of production (products with extremely low (or non) material content and the surge of the "new economy". The idea of dematerialization of production describes a twofold process in which: a) added knowledge is the principal component in the value of the final product, while the relative costs of physical materials decline constantly, and b) the most dynamic goods and services in the economy are those that transmit (as is the case with goods connected to the culture industry or finances) or condense/incorporate information (as is the case of medicines or genetically modified seeds).

The new economy related to "dematerialized" products is dominant in the telecommunications, audiovisual, biotechnology, nanotechnology and pharmaceutical industries. The principal characteristic of these companies is that they are dependent on permanent technological innovation, which transforms knowledge into products and services. The market value of enterprises based on research or innovative products, is not only related to their current income levels, but based on the projections of their potential for future sales if the product/service they invent is adopted by the market. The new economy has transformed a considerable part of financial investment into venture capital, as it is carried out under high-risk conditions in which expected potential gains may never materialize.

4) The deepening of the process of individualization, in the sense that there has been a reduction of outside references in standards and values of social conduct. Individuals are no longer guided by traditional values, norms, institutions, and ideologies of modernity (such as patriotism, parties, work, family), bringing about a new form of reflexive individualism in which people must constantly negotiate social relations (for example with spouses, sons, daughters, and colleagues). By inserting the reflexive individual in a world of global information and increasing contacts with diverse social networks, information technology enhances individualism.

5) The globalization process and loss of symbolic significance of the nation, is a manifold process. New transnational agents have been proliferating since the 1960's, when multinational companies began acting on an international scale according to a strategic vision that is not delimited by national borders. In the past decades the number of these transnational agents has multiplied due to the growing internationalization of various groups, including financial capital, scientific and technological systems, religious groups, non-governmental organizations, and criminal and terrorist organizations. The processes of internationalization of financial flow, of international commerce, and of patent regimes, have limited the breadth of action available to governments. Meanwhile, the Internet facilitated the globalization of social and cultural interactions, limiting state control over sources of information and restricting the ability to limit cultural systems to those compatible with national characteristics. The consequence is the acceleration of the formation of worldwide public opinion movements.

Despite this, the state continues to be the principal actor in national and international politics and the nation the main locus of life chances for most citizens. If the globalization of societies has limited governmental freedom of action, by disseminating a cosmopolitan agenda and global standards of consumption it has also increased the expectation that governments will ensure access to the new rights and goods.

## *3.4 What are the main social consequences of the information revolution?*

**KEYWORDS:**

SOCIETY, CULTURE, SPACE-TIME PERCEPTION, HYPERTEXT

Information technology has accelerated the following processes and had specific impacts on knowledge and culture. The first, about which there is some consensus among researchers, is the unification of the perception of space-time at least in relation to all of the dimensions that are based on the flow of information (in the form of text, voice, or image). In human experience, the limited reach of the bodily senses require that individuals go from one place to another to reach another individual or place, determining the sense of distance as related to time. Mechanisms for sending information such as drums, smoke signals, mail, telegraph, telephone, and television, were the means developed by humans to transmit information without going from one place to another. These tools brought a new dimension to the relationship between time and distance.

Now that voice, text, and images can be instantly transmitted, the association of space with time is disappearing, at least with relation to things that can be transmitted digitally. The feeling of a global village is accelerated with each new communication technology. With the introduction of transmission of television images via satellite, televised events take place for the whole world at the same instant regardless of space or time. The Internet brought this revolution to a new level allowing an individual in any place on earth to be in immediate contact via a choice of voice, text, or images, with any other person on the planet. At the same time, it brought a large part of the collection of human knowledge and culture (at least those parts that can be transformed into digital format) to cyberspace, making it available for any user in any place.

**The unification of space and time does not mean that temporality has disappeared. On the contrary, it represents the contraction, acceleration, and increased value of time brought about by the disappearance of space barriers.**

A second aspect, about which there is debate and sometimes confusion regards the concept of **virtual reality**, defined as a **set of images and sensations produced electronically**. Virtual reality is often contrasted with "real" reality, as though the virtual world were less real or authentic than the world of sensations or the world as we experienced it before these new technologies. In general this is a romantic vision of the past, of sensory experience, and social life. The world of humans was always "virtual". Human beings relate to their world through individual imagination structured around cultures, a set of abstract symbols that determine how things transmitted by the senses are perceived, understood, interpreted, and evaluated. Be it by way of the Bible, the Koran, or a science book, the only way to get beyond the finiteness of individual experience is through the world of meanings that organize our perception.

The third impact of the Internet is perhaps the deepest and still in its early infancy. It is the transformation of the human universe by the growing integration between machines and humans. This is an area where speculation abounds and of which there are two main schools of thought. For some, the computer has the potential to mirror the human mind, allowing integration between the two in the future. For others, the distance between the human mind and artificial intelligence is unsurpassable because the human mind cannot be dissociated from the biological and cultural support that allows it to function.[5]

An ample bibliography already exists concerning the consequences of hypertext on intelligence and perception. Hypertext allows the development of reading written material in permanent connection and association to other texts, thereby allowing instant access from one text to another in a continuing spiral. It is different from "classic" text reading: from beginning to end. Some authors claim that the capacity to connect large amounts of information associated with diverse contents and networked material could cause losses in the intellectual culture of the "the age of books", with its emphasis on deep reflection and conceptual development carried out in large part by individuals making isolated efforts. Others emphasize that the intellectual activity associated to hypertext takes place with more awareness of the collective character of all works, is less individualist, and more fluid.[6]

## 3.5 What are the limits of the information revolution?

**KEYWORDS:**

SOCIAL CHANGE, FACE-TO-FACE RELATIONS, STRATIFICATION, POWER RELATIONS, SOCIAL VALUES

Since the Internet evolved into the most commonly used means of distance communication (substituting the post and to a degree the telephone), naturally it's presence has permeated all social, economic, and political relations. But this does not mean that the Internet is capable by itself of changing these relations. If the Internet, as we have shown, deepens existing trends in contemporary society, nothing so far, indicates that it is a factor in radical transformations of social structures, systems of stratification, or the norms and values of society.

Electronic networks do not substitute face-to-face relations, which **continue to be the principal source of trust in**

**human relations**. The possibilities that the Internet opens intensify interchanges and diversify social networks but with the exception of marginal cases, generally of adolescents approaching adulthood who encounter an alternative world on the Internet, this has not been sufficient to substitute or modify in a dramatic way the social ties that are established by direct coexistence.

## *3.6 What are the main social applications of the Internet?*

**KEYWORDS:**
E-MAIL, INFORMATION, SCIENCE AND TECHNOLOGY, PRODUCTION, E-EMPLOYMENT, E-CULTURE, E-GOVERNMENT, E-HEALTH, E-CRIME

Being an all pervasive technology in order to understand the impact of the Internet on social inequality we need to map some of its most common applications.

**E-Mail**: Through the vehicle of e-mail the Internet offers a mechanism for sending messages and documents instantly causing the postal mail (as well as of the telegraph, fax, and telex) to practically disappear as a means of transmission of text and even to a certain extent substituting telephone calls. The Internet has changed the rhythm of communication requiring greater speed and creating the expectation of immediate responses. However, the speed of human emotional and intellectual processing, based on evolution, is not the speed of light, creating new sources of stress. E-mail is the most common use of the Internet, and increasingly, having an e-mail address is viewed as the equivalent of a residential address, a way that a person can be "located". The lack of an e-mail address will cause social isolation. In the future an e-mail address will be a basic condition of citizenship.

**Information**: The Internet has facilitated the process of accessing information enormously, a process which by traditional methods required great investments of time, energy, and resources. The Internet does not just store computerized cultural production; it is also a way of making the material publicly available. The Internet allows access to a growing collection of text, images, and sounds to which the public would not have access if material reproduction were necessary. The growth in the amount of material available through Web sites is exponential causing users to depend increasingly on research mechanisms to locate information of interest. While on the one hand the fact that anyone can put content on the Internet represents a form of democratization of information, on the other hand it causes dependence on search engines that have the capacity to influence the priority level of texts for users.

It is possible to distinguish between the use of the Internet to get information and knowledge, involving material with high and low informational content. Low informational content material refers to facts that do not require any special intellectual training to understand and comprehend and which are depleted after serving their immediate function. For example the name of a street, a pornographic image, a bank transaction, or online shopping, are all low-content information. High informational content material depends on the analytical ability of the user and has an impact on his or her further competences and intellectual abilities. As we will see, the prior intellectual ability of the user is a determining factor in the transformation of the Internet into a tool of cultural empowerment and social creativity.

**Science and Technology**: in addition to facilitating access to databases, virtual libraries, and all kinds of information, the Internet has enhanced two traditional characteristics of the fields of scientific research and technology: the functioning of networks and the international character of interaction. These professional sectors are more open to the Internet and its possibilities for restructuring communication, and are also among the most influential to the extent that the strengthening of international networks related to the interests of each researcher has contributed to the weakening of immediate social ties based on departments and faculty life.

The ability to circulate scientific work on the Internet allows new possibilities for scientific publication in electronic periodicals without printing costs. This has contributed to academic debate concerning the regulation of these publications (whether they should follow the same editorial norms as printed publications) as well as disputes concerning intellectual rights.

**Production**: as discussed earlier, information technology did not globalize the economy itself, but it accelerated communication between and within companies enormously, independent of the location of any given employee, increasing the speed and quantity of informational interchanges. The Internet allows companies to keep track of inventories, market trends, and relations with providers and clients online, reducing the time of the production, distribution, and consumption cycle.

The impact of information technology on the productive system is varied. In the first place it has created an enormous market for information technology products, from communication infrastructure to computers, equipment, and software. Second, it opens up the possibility of new products and services that can be transmitted via the Internet. Third it has allowed revolutionary changes in systems of knowledge management and communication within companies. Fourth the different forms of e-commerce (the principal forms are B2B - business to business, B2C - business to consumer, C2C - consumer to consumer, and B2G - business to government) have revolutionized selling and buying. Fifth, the electronic auctions of B2B, B2G, and C2C have reduced transaction costs enormously while B2C and B2G have modified logistics and the supply chain between and among businesses while increasing the speed and reducing the costs of transactions. Finally, services that were previously carried out by employees serving clients can be transferred to the client, as is the case of automatic teller machines and Internet banking, or secretarial services which are now in large part carried out by each employee, or with the reduced importance of the sales staff in commerce between businesses.

The tendency to contract time has had a particular impact on the financial sector, one of the sectors that due to strictly informational nature of money, has come closest to the elimination of time. As time can never be eliminated, the financial sector has always been characterized by the fact that many lucrative opportunities depend on the ability to arrive first. Today this translates to an advantage that is counted sometimes in terms of seconds.

**E-Employment**: The Internet allows all information to be encountered in virtual space. Because all information can be accessed in virtual space, the necessity to use physical space has diminished and contact between the employee and the employer has become independent of their location, allowing increased productivity and making the structure of businesses more flexible. Transmission of messages by Internet has changed labor practices in the service sectors where information is circulated via e-mail permitting more agile communication, documentation, and control.

If the Internet allows companies to be reached any place, when combined with cellular telephones, it also allows em-ployees to be reached wherever they are. The consequences of this have been calamitous, increasing the amount of work that is performed outside normal work hours and in practice, destroying the notion of work hours, weekends, vacations, and the distinction between work and the private sphere. The rhythm of electronic communication enters into conflict with the biological and emotional rhythms of people and this conflict leads to growth of social problems and ills, whose current symptoms are the epidemic of stress and depression caused by the difficulty of keeping up with the rhythm of things. Sooner or later it will result in demands for new regulations in the world of work. The worker's rights will need to include the **right to remain unplugged** outside of work hours. If we do not move in this direction in the near future, humanity will have to reinvent one of the principal contributions of the Bible: the right to a day of rest.

**E-Culture**: A growing part of the potentially digital aspects of humanity's cultural heritage is available on the Internet. It's already possible to visit a large number of museums, virtual historical archives, and virtual libraries. These collections comprise a large part of the great works of literature, at least those works that are in the public domain. These materials can be obtained on the Internet usually without cost. In the future every new musical work, film, and literary work will be available via the Internet.

**E-Government**: the impact of the Internet on political life can be divided into three levels: **e-governance** refers to the use of the Internet for increasing efficacy, quality, efficiency, transparency, and enforcement of the actions and services of the government and public institutions; **e-government** includes a set of new instruments that allow greater and different types of citizen participation in government decisions; e-politics refers to the impact of the Internet on the social structure and the political organization of society.

E-governance allows the use of the Internet for: 1) publicizing all activities of the government including budgets and public spending, allowing greater transparency and public monitoring; 2) improving the quality of administrative services by increasing their speed and outreach; 3) offering services online, including government documents, health and education service requests, bill payments, and tax declarations; 4) the electronic transmission of public bids and auctions.

E-government includes electronic voting, the possibility of interacting with public institutions, and regulating activity associated with the Internet – development of legislation concerning commercial activities, security and individual privacy rights – as well as all measures designed to ensure universal access to the Internet.

**E-Health**: The Internet has facilitated work in the areas of monitoring and controlling epidemics, reorganizing healthcare systems and patient relations, and allowing access to medical information by laymen (a trend that is frequently criticized for producing erroneous or counterproductive information). In the area of health information technology is particularly promising despite its limited impact at the moment. There have already been several successful experiments in tele-medicine including diagnosis, distance surgery, medical teleconferences, and tele-monitoring. The majority of these experiments are still in the pilot stages.

One service that has been developed in many advanced countries is the so-called health card, a card that allows access to patient medical histories regardless of location. The health card facilitates remote health services and medical research. If the confidentiality of these databases is not protected it is possible that insurance companies and employers in possession of this information could develop discriminatory insurance and employment policies.

**E-Crime, E-Terrorism, and E-War**: Finally, we can't forget the potential of information technology to be used for falsification, theft, and destruction by criminal and terrorists networks and by a new type of criminal, who specializes in breaking the security systems of networks and sites for destructive goals. Crime and terror have been, so far, much more effective at making use of new information technologies than most of the security systems of the national states. This is particularly dramatic in less developed countries.

*3.7 What are consequences of Internet on education?*

**KEYWORDS:**
PEDAGOGIC SOFTWARE, SCHOOLS, UNIVERSITY, TEACHERS, LEARNING, TECHNOLOGY EDUCATION

E-Education, i.e., the capacity to analyze, bring together, and make use of information is a central component of professional competence for the majority of economic activities in the contemporary world. In principle, the Internet and Education seem to mutually reinforce one another but in practice the relationship between them is quite complex. The intersection of education and information technology has two axes: the transmission of specific educational content, and education oriented to further development of the capacity to use information technology independently.

The use of the Internet to develop specific competence or knowledge (language education, extension courses, professional courses, and courses in diverse areas including higher education) is widespread today. A growing number of companies, and practically every university in the developed world and many developing countries have multimedia production facilities and/or distance education courses. Educational CDs, which in many cases require only computers and not access to the Internet, represent an important segment of the education markets.

Although there is still very little long-term comparative data concerning the effects of the Internet on adult education, several international and governmental institutions have compiled reports attempting to evaluate the impact of the Internet on education. They indicate positive results in the area of second language instruction, training for the business sector, higher education, and teacher education. Private industry and universities have been functioning in all of these well-established markets, which have been little affected by the recent crisis in the new economy.

At school level, aside from special cases such as children with special needs, the impact of the Internet on education appears to be ambiguous. There is a shortage of long-term and comparative studies for clearly identifying the contribution of the Internet in elementary schools. Case studies indicate that teacher training continues to be a fundamental element in the educational system and that the Internet can be used as a complement, but not as a substitute, for the functioning of the teacher. The principal differential in terms of individual performance in school, aside from social context and family background, continues to be the teacher's qualification level. Through the words (and emotions) of teachers, children develop intellectual instruments that allow them to advance their reasoning capabilities and analytical autonomy that are so fundamental in the Internet age, where the availability of an unlimited quantity of information can be practically as paralyzing as the lack of information.

In addition to personal relationship between teacher and student, there is no substitute (at least not in the near future)

for paper and pencil, both because of its importance in the development of writing skills and for the value of paper as the best means of storing information and work compiled by students and making this work available for teacher and parent review.

New technology tends to transform the role of the teacher by subverting his or her function. If it offers a great potential for supporting classroom activities, when used to substitute the role of the teacher it limits the creative application of his or her pedagogic experience and interaction with students. In practice, some interactive educational software excludes teachers from their pedagogical functions.

Since the introduction of the Internet in education is still experimental, it should be carried out gradually, backed by the experiences of pilot programs and **as part of the (much needed in most countries) general reorganization of the teaching system**. For example the three-dimensional computer images are an excellent tool for facilitating comprehension of things like the human body, the subatomic world, or geology. But using these tools to develop children's complex reasoning will require more advancement and redefinition of the role of the teacher and curricula. This type of instructional practice is still its initial stages and the instruments that are used still need improving. Since multimedia products tend to standardize education, they tend to move away from the requirements for individualization, an adaptation to the necessities of each student, especially at the school level.

At the secondary and university level, excessive emphasis on the importance of the Internet as a source of information and ideas can have damaging consequences. Research indicates students have substituted reading and writing efforts with Internet searches for texts that can be adapted to meet their assignments. Rather than being used as a starting point Internet searches have been transformed into arrival points leaving behind the practices of sustained reading and reflection. The excessive emphasis on the computer screen and on multimedia educational tools risk compromising the pedagogic necessity of developing the intellectual discipline needed for reading a book and or the patience necessary to develop creative ideas.

The indiscriminate introduction of computers and Internet can have negative effects on education especially when teachers lack adequate training in computer and Internet use. Massive investments in teacher training will be necessary to avoid gaps between the knowledge of teachers and students in relation to technology use.

**Education supported by information technology** should not be confused with **information technology education**, an urgent necessity in all school systems. Information technology education requires the creation of mandatory courses designed to prepare students in the use and evaluation of information technology instruments, from their technical basics and uses to considerations of the challenges they create for society. The introduction of computers as teaching instruments should be preceded by teacher education programs designed not only to prepare teachers in purely operational terms, but also to offer them a more general understanding of the environment in which computers and the Internet function as research tools that can advance student ability to pose questions rather than simply finding means and not ends.

## *3.8 What are the consequences of the Internet on politics?*

**KEYWORDS:**
DEMOCRACY, SOCIAL NETWORKS, POWER, PYRAMIDS

An analysis of the impact of the Internet on politics must not overlook the fact that there is a historical tendency to associ-ate the way that policy is made with the dominant means of communication. Mass society is often associated with Radio and in some cases even ascribed to it. In the same way that television has promoted the "spectacle society". Today Internet is related to a new way of making policy through the strengthening of civil society networks disassociated from, or marginally connected to, the state.

Though past results do not permit predictions for the future, we must not forget that many socialists regarded the radio era as the advent of a new era of popular participation. We lack sufficient evidence for confirming theories regarding the impact of the Internet on politics. Currently there co-exist two opposite interpretations of its consequences: some analysts believe that we are about to experience a radical social transformation from representative democracy to referendum democracy while for others, new forms of virtual socialization could destroy the basis for interaction that allow the construction of public space and increases the capacity of control over the population by the state or by political marketing companies.[7]

The interactive and open character of the Internet have caused many authors to view the Internet as a source for a new paradigm of social organization in which the central category is the **social network**, a system of communication comprised of interconnected nodes which are fluid and constantly changing form. In this model, each social actor participates in different networks each one depending on one another within relations that may be asymmetric or hierarchical but within which all parts are interdependent without a defined center. The network would subvert the hierarchical and rigid social structures of industrial capitalism, which in contrast are characterized by vertical systems of communication with well-defined structures of power.

In the 20th century the dominant metaphor for describing society was that of a social structure taking the form of a **pyramid**. From the pyramidal perspective, society is organized with a small top representing the most rich and powerful with workers at the base, while some sectors of the population are in the middle. In another metaphor of social organization societies had a **center** and a periphery; companies divided themselves between employers and employees, while the political structures were based on dominant and dominated groups. In this model, interaction and communication between the different levels was underplayed. However if the mere existence of the state implies a power structure, social participation was always present in political parties, social movements, and public opinion. If companies were based on authoritarian power systems, workers responded with unions or factory commissions. Finally if the distribution of resources stratified society, social mobility presented a more flexible reality. Even highly centralized totalitarian states depended on networks for keeping informed and controlling society.

In contemporaries societies it has become obvious that the unilateral nature of the pyramid metaphor is inadequate. The question is whether the use of the network metaphor should cause us to completely abandon the idea of power structures or society as a pyramid. We believe that both concepts are only partial explanations: information technology has affected the techniques of power transforming traditional systems of organization and the functioning of decision-making centers without eliminating them.

Rather than viewing the metaphor of the network as a radical break between the present and the past, we should consider how networks have always been a central part of human society. In fact the importance of communication and information has been a central theme concerning the nature of capitalism since the work of Adam Smith. Not only the markets but also democracy has all the characteristics of a network, where participation in the circulation of information is central for its functioning.

The danger of a unilateral emphasis on the role of networks is that we move from a partial metaphor of structures and pyramids to another equally insufficient metaphor. Virtual networks multiply and change the functioning of vertical organizations, which increasingly make wider use of networks, but they come a long way from making those structures disappear. In the same way that the unification of space and time has not eliminated time, the creation of virtual networks has not eliminated the material nature of the world and the importance of the centers of political and economic decision-making and control of power and economic resources. The principal source of technological innovation in telecommunications continues to be investments in research connected to the military sector, a highly centralized structure.

Networks and structures have always been interlinked. National governments have always known about the importance of the means of communication for unifying and controlling national space. The centers of power concentrate political, cultural, and economic resources imposing unequal distribution of decisions and communication flows. If the Internet indeed has the effect of weakening territorial trends, by allowing interchanges on a global scale, territorial spaces continue to be important not despite but because of their material nature, their human resources and their infrastructures.

However the illusion of a world of participatory global networks functioning apart from the government and corporate sector power structures of global society not only offers a partial vision of social reality, but is also problematic because it causes us to overlook new problems in global society and to abandon the dialogue with organizations (states and large companies) that continue to be decisive actors in con-temporary society. The great intellectual challenge at the beginning of this century is to invent new forms of interaction between pyramids and networks to increase the democratic potential of new technologies.

## 3.9 How do networks and centralized power interrelate?

**KEYWORDS:**
PARTICIPATION, COPYRIGHTS, FREE SOFTWARE, PRIVACY, GLOBAL VILLAGE, FREEDOM

The optimistic vision includes disparate views. For some the Internet allows increased citizen participation in government decisions by way of a system of ongoing consultation allowing day-to-day referendums on diverse themes. For others, the Internet enhances democracy radically, creating a new public space in which civil society organizes itself separately from the state. The pessimistic perspective includes those who consider the Internet as a threat to face-to-face relations, the only source of communication capable of generating solid and stable groups with historic memory (rather than the a-temporal world of the Internet), and capable of sustaining public life and constant political action. By creating a world of virtual relations the Internet facilitates the growing control of governments and of corporations over citizens, destroying privacy and liberty.

The diverse positions about the impact of the Internet indicate real tendencies that depending on the outcomes of social conflict could someday become dominant. In practice both the strengthening of democratic life and the weakening of privacy and freedom through the control of information can be encountered today. Databases that centralize information from genetic codes to medical histories, laws that require service providers to keep copies of all e-mails, cameras that film activities in workplaces, streets, and stores, tracking systems for Internet users, credit cards that record details on all purchases, electronic toll booths, cellular telephones with cameras and GPS systems, new systems of biometric recognition, and in the future microchip implants with medical or other functions, converge, destroying the notion of privacy and together comprise an enormous potential for social control and the destruction of freedom.

The global village runs the risk of reproducing the aspects of traditional villages that made them into places of control and social oppression and where anonymity and the feeling of freedom were impossible. The growing impossibility of lying (as an individual choice) has a destructive potential for human sociability, as we know it. Though lying can be used to hide crimes, it is also a defense mechanism for the underdog and the oppressed, and a fundamental recourse of human freedom. The impact of the Internet is in fact **bi-directional**. If on the one hand it expands the possibilities for action, for worldwide public opinion, and for activism among decentralized social movements it also allows new forms of social control as well as facilitating international networks of organized crime and terrorist groups.

The dependency of society on networks of electronic communication for the proper functioning of practically any service creates an enormous risk of paralysis and destruction on a global scale in case of successful attack on the system. Processes that make humanity vulnerable have always been a part of the interactions between the diverse groups of people made possible by the encounters of societies. When people lived in isolation they did not have access to others' technological and social innovations, but they were also protected from diseases, epidemics, and new problems that are parts of the homogenization of productive, social, and political systems. With globalization, epidemics travel by airplane and an electronic virus at the speed of light. The homogenization of crops is now on a worldwide scale and advances in medicine save life but can have a neutralizing effect on bio-diversity and the natural selection mechanisms of the species, impacting on all of humanity and the planet rather than local people and ecosystems.

New technologies have increased individual communication and access to information but also have built a fragile system of life, dependent on technology which is vulnerable to attacks and whose existence could lead to totalitarian practices of social control. It has also changed the art of war, thanks to new satellite systems and remotely guided missiles that combine, with increasing efficacy, information, communication, and destructive power. Perhaps there is a tendency in human history for every new technological instrument that increases the capacity to control nature and society and which simultaneously promise improvements in the quality of life, renewing the hope of a better world, to at the same time increase the potential for destroying the environment and society.

## 3.10 Why have copyrights become a central political issue for Internet activists?

**KEYWORDS:**
PARTICIPATION, SOCIAL MOVEMENTS, INTELLECTUAL PROPERTY

Internet networks (represented by citizens and consumers) and pyramids (represented by the states and by corporations) tend to confront one another. While large companies connected to the information technology industry promote the broadest interpretations of copyrights in order to transform every piece of information and artistic works in private property, individuals and voluntary groups have developed free software and free access sites and alternative copyrights license models, like Creative Commons. Government security services try to control the communication and information of each citizen (and in dictatorships explicitly censoring and limiting the access to the Internet) individuals and organized groups work to limit these powers and strengthen privacy rights.

To the extent that scientific knowledge, information, and culture can be transmitted by the Internet, the most appropriate forms of social regulation of the system have become the basis of a worldwide confrontation. Information, knowledge, and culture can be seen as merchandise and sources of profit or as public goods that should benefit from state intervention to ensure effective universal access for the population. The two main issues are the regulation of copyrights and the governance of the Internet.

Public debate over what can be patentable became a central issue for civil society already in the 1980's, mainly related to biotechnology products like genetic engineered seeds and later human genes. Historically the concept of the patent was founded on the distinction between invention and discovery. While the former could be patented, as it was associated with the creation of something new that does not exist in nature or in the public domain, discovery refers to knowledge of something pre-existing in nature or society, as is the case with scientific knowledge, which cannot be patented. The objective of patenting was to ensure that new knowledge would not be transformed into industrial secrets. As an incentive for inventors to put their inventions in the public domain they were insured a monopoly of use or royalties from third parties for a certain limited period of time. With biotechnology, the separation between discovery and invention was called into question, with attempts to patent knowledge concerning the utility or function of certain genes, giving the patent holders the rights to charge royalties to anyone using this knowledge to develop new products. Thus science was colonized by industry and the knowledge that it produced started to lose its universality and free communication character which is among its chief historical characteristics.

In the intellectual field, production copyrights were originally used to protect editors and later authors of all types of artistic and intellectual works. In the European tradition, copyrights protect the author and in the United States copyrights can be transferred in their entirety to companies. Recently copyrights of all literary and artistic property were standardized worldwide from 50 to 70 years after author death and 95 years after publication or 125 years after creation in cases where the rights of a work have been acquired by a company.

Copyright law has always accepted fair use clauses, including the reproduction of work for educational research and for personal use. With the widespread use of photocopiers the meaning of fair use was already being questioned by editors. Later, with the introduction of Internet, the copyright problem become central because of the possibility of placing almost any written, musical, or visual work on a Web site at practically no cost short-circuiting the owners of copyrights. Various sites specialized in making music and later films recently released available at no cost on the Internet. Recording companies responded by demanding the closure these sites in court. Though the companies won in the courtroom, the near impossibility of eliminating these sites has caused the decline of these companies and sooner or later will force them to redefine their business model.

The new trends in copyrights have allowed the transformation of activities that were traditionally considered crafts such as teaching classes, organizing courses, conferences, and developing pedagogic material, into potentially patent-able products. Institutions began to regard their staffs as producers of patentable material, transforming work that used to belong to the nonprofit category into income generating activity and by so doing they have modified the scientific ethos. In the United States, in particular, demands for copyright payments by third party users of any innovation, text, or image, even marginal or tangential uses, has become a mania with paralyzing impact on creativity.[8]

The Internet has raised questions regarding the definition of fair use in the cyberspace context. Should permission or payment be required for non-commercial use of information and material made available on the Internet to anyone with access to a Web browser or search engine? Should the browser or search engine receive payment? Internet companies and holders of copyrights expect than in the future they will be able to require payment from each user for any downloaded information. To gain control over Internet users, many pressure the hardware industry to include mechanisms that control and monitor each user action.

If solutions to the question of payments for sites and search engines center on individualized solutions including control mechanisms that take away the public domain character of the Internet, companies will gain unprecedented access to private information. An open national and international public discussion should seek to identify answers that allow payments for content producers while ensuring the open and public service character of the Internet and preserving personal privacy. One proposed solution is a fee-based system, collecting from users and distributing among visited sites according to a publicly controlled system.

## *3.11 What is the open source movement?*

The creators of the Internet left the source code of their work in the public domain allowing any person to develop compatible programs without paying for copyrights. Since the beginning of the Internet a group comprised primarily of researchers and hackers have worked to keep source codes for computer programs in the public domain. This movement seeks to confront the growing oligopolisation of source codes within the software industry such as the best-known Microsoft case. These movements created the Open source Standard, certifying that the source code is available without cost to individual users and Creative Commons, as an alternative intellectual property system.

The most important product of the open source movement is the Linux operating system. The use of Linux is widespread within large companies, institutions, universities, and governments. These groups generally use large computers with sufficient human resources for providing technical support to users. For individuals and small businesses without technical knowledge, commercial software has the advantage of easy installation and technical support. Although its use continues to be small, there are increasing number of companies that specialize in providing technical support for open source programs.

Discussion concerning open source software often revolves around economic themes, particularly in developing countries where open source software use is viewed as a way of reducing expenses. The existence of open source software has, in fact, pressured the industry to reduce prices. But the central question raised by the open source movement is fun-damentally political. The original architecture of the Internet facilitated communication rather than identification of the user and the content being transmitted. Both companies and governments have since developed tools aimed at identifying users and their movements. For companies, monitoring users each time they access the Internet is the only way to ensure payments of copyrights while also offering information regarding the consumer profile of Internet users. For governments, access to and storage of information transmitted in cyberspace is seen as a necessary component of vigilance over activities that could affect national security, causing many countries to restrict or prohibit the use of cryptographic systems by private parties, to regulate the commercialization of advanced deciphering systems, and to oblige service providers to store all e-mail for a certain period of time.

The existence and ongoing development of alternative open source programs is among the fundamental conditions, together with copyright laws, for limiting the capacity of the corporate sector, to subordinate the Internet to the logic of their interests. For the large majority of information technology users, commercial programs and hardware are black boxes. Most users lack a notion of the information that they transmit when they use the Internet. More public debate is needed to assess the public liberties threaten by the power of the state and the corporate sectors to interfere, monitor, extract and make use information from Internet navigators and email users and define the rights and ways of private companies to collect returns on their investments and of the state's concern with public security.

## *3.12 What are the Internet governance challenges?*

**KEYWORDS:**

INTERNET GOVERNANCE, PUBLIC SPACE, INTERNATIONAL REGULATION

Cyberspace represents one of the great challenges to the new forms of international governance. Besides the problem of copyrights, previously discussed, the regulation of cyberspace involves two major issues: 1) the governance institutions of the Internet; and, 2) the regulation of cyberspace as a public space.

The first issue attracts most of the attention of social activists, and the main focus is the governance structure of the Internet (the ICANN - Internet Corporation for Names and Numbers – known widely as ICANN –). The ICANN was created as a non - profit organization, by the USA government Department of Commerce in 1998, and has as its main responsibility the allocation and management of IP addresses, domain names and the root server system. The main political issues around the ICANN are related to its scope and the role that it should play in it the United Nations, national governments and civil society.

Cyberspace has become a central part of the public sphere and as such poses the problem of regulation. The challenge we face now is how to prevent this tool from being colonized by antidemocratic groups, i.e., that the potential exchange and debate of ideas not be thwarted by the dynamics that the new medium itself may generate. There are good reasons to be suspicious about states trying to control the contents and messages that circulate on the Internet. Notwithstanding, given the fact that the Internet is becoming the privileged medium for public debate, it is now necessary to analyze and evaluate the specific characteristics of communication via the Internet, so as to create regulating mechanisms, which as far as possible do not depend on state interventionism.

# 4. DIGITAL DIVIDES

## *4.1 What is the Digital Divide?*

### KEYWORDS:
DIGITAL DIVIDE, SOCIAL INEQUALITY, ACCESS TO COMMUNICATION
TECHNOLOGIES

The term digital divide refers to the unequal access to a diverse collection of communication tools such as radio, telephone, television, and the Internet. Although we will be focusing on access to and uses of the Internet, the digital divide cannot be disassociated from access to other communication technologies and in fact each of them produces a specific divide. Therefore rather than a digital divide (related to only one given communication artifact or an abstract average use of all the communication vehicles) it would be more precise to speak of digital divides. This chapter will focus on the digital divide produced by the Internet.

There is a strong correlation between the digital divides and other forms of social inequality. Generally the highest levels of digital exclusion are found in the lowest income sectors. Communication divides do not manifest themselves solely only on the basis of access to material goods such as radio, telephone, television, and Internet. Each user's intellectual and professional capacity to make the most of each of these communication and information technologies is as important as access itself.

Although most of the literature on digital inclusion, especially reports produced by international agencies on developing countries emphasize the potential of ICTs to reduce poverty and social inequality, the social dynamic is quite the reverse: the introduction of new ICT's generally increases, at least at the beginning, social exclusion and inequality. Univer-salization of access can limit the damage, and sometimes offer new opportunities, from as regards social inequality.

## 4.2 Why digital exclusion can't be separated from other forms of inequality?

**KEYWORDS:**
POVERTY, QUALITATIVE STUDIES, QUANTITATIVE STUDIES

Poverty is not an isolated phenomenon. How poverty is defined and perceived depends on a given level of cultural/economic/technological/political development in each society. The introduction of a new product that becomes a condition of 'civilized' life (be it a telephone, electricity, a refrigerator, radio or TV) raises the minimum standard by which one is defined as poor. Richer sectors of society are generally the first to have access to new products, and it takes a long time before these products are made available to the poor – if at all. Therefore, the introduction of new 'essential' products increases inequality.

Moreover, since richer sectors of society are the first to have access to new products, they have the benefit of initial competitive advantage in using and mastering them. At the same time, those who are excluded face new, or greater, disadvantages. In both cases, new ICT products increase, in principle, poverty and social exclusion. The main aim of digital inclusion policies therefore is to diminish the negative impact of new ICTs on wealth distribution and life opportunities.

Most of existing in-depth studies on digital exclusion focus on small communities or local experiences and do not integrate studies based on quantitative data. On the other hand, statistical studies – in particular those on developing countries – have as a central and generally unique parameter: the division between those who have and those who do not have ac-cess to computers and to the Internet. Although important, this measurement is insufficient to understand the broader social dynamics and to define policies which make access universal, because of three important factors:

a) They do not identify the quality of access, whether in terms of connection speed or cost/access time available, in particular for the poorest groups of the population.

b) When quantitative studies do distinguish between socio-economic strata, they use possession of a computer in the home as the basic criterion for access, while in fact tele-center and cybercafés, workplace and family or friends with access are as important for the poorest sector of the population.[9] Most of these studies do not give information on different types of uses, and its relevance for users.

c) Digital exclusion is not only about access to the Internet versus those who do not have access – of those who are included versus those who are excluded.[10] While there is a real polarity, it sometimes masks the multiple aspects of digital exclusion, which are related not only to access but also to appropriation, the capacity of making sense, interpreting and making use of the information available on the Internet.

The digital divide represents a dimension of social inequality: it measures the relative level of access to products, services, and benefits of new information and communication technologies between different segments of the population. The digital divide also addresses another subject associated with social inequality that cannot be confused with the digital divide itself, that is, information technology as **a tool in the fight against poverty**. In situations of economic growth it is possible to reduce poverty indicators (the size of the population below a set poverty line), while simultaneously increasing social inequality. Thus the fights against inequality and poverty have some commonalities but are not synonymous.

As already mentioned, the initial social impact of the Internet did increase social inequality because it reached first the wealthiest sectors of the population. Thus, the fight against the digital divide is not so much a fight to diminish social inequality in itself as it is an effort to prevent inequality from increasing because of the advantages that those groups of the population with more resources and education enjoy as a result of exclusive access to this information technology.

## 4.3 What are the different aspects/dimensions of access to Internet?

**KEYWORDS:**

PHYSICAL INFRASTRUCTURE, TRAINING, INTELLECTUAL CAPABILITIES, SOCIAL INSERTION, CONTENT PRODUCTION

The digital divide depends on five factors that determine the level of equality of access to information technology systems: 1) the existence of physical **infrastructure** for transmission; 2) the availability of **connection equipment** such as a computer, modem, and access line; 3) **training** in the use of the computers and the Internet; 4) **intellectual capabilities and the social insertion** of users (this is the product of the educational and intellectual level, profession and the social network that determines the effective use of information and the necessities of Internet communication; 5) the **production and use of specific contents** adapted to the needs of the diverse segments of the population. While the first two criteria refer to passive dimensions of Internet access, the last three dimensions define the potential of its active appropriation.

The distinction, between the different levels of access and use, is basic to development of methodologies for evaluating, accompanying, and acting in the fight against the digital divide. Public programs aimed at universal communication services focus primarily on the first and second levels of physical infrastructures and connection equipment.

Access infrastructures are comprised of transmission systems that can function by way of telephone, satellite, radio, cable television, electricity wires and cellular phones. In the future it will be possible to access the Internet by way of interactive digital television. Internet connections can be by fixed phone dial connexion, or a variety of broadband systems of access, which determine the speed of information transfer. The availability of individual access is dependent upon the existence of local providers for these services.

The universalization of access infrastructures is a process that is practically complete in advanced countries, though there are still some isolated regions where broadband is still not available. In developing countries, on the other hand, the universalization of infrastructures is still a central problem, particularly in rural areas and remote villages. In the developing world, broadband is generally only available in large and some medium-sized cities. In the majority of developing countries, the Internet is concentrated primarily in large cities.

The most common equipment for accessing the Internet for the poor in developing countries is still a computer with a modem and a telephone line with a dialup access to a service provider. The main mechanisms of individual access are: home, work, school and public or private tele-centers. In low-income sectors without equipment or access services, access to the Internet depends on **collective access points** such as school, work, or tele-centers (non-profit collective access points) and cyber-cafes (private profit oriented access points).

The bibliography on the digital divide is generally consistent in defining two main factors that determine Internet access levels, given the existence of communication infrastructure. They are personal income and educational level. Given the same income level, people with higher education levels are more likely to have access to the Internet. With the exception of some particular regions, there is relative equality of Internet access between men and women. The unequal impact on different racial and ethnic groups tends to be consistent with inequalities in income and education, with the exception of some situations where unequal access is aggravated by language differences between ethnic groups. The penetration of the Internet in developing countries is also associated with the level of urbanization. The digital divide, particularly in the developing countries, is aggravated dramatically in rural regions. In general, higher levels of urban concentration correspond to a higher numbers of users.

The digital divide has a strong age component that is more pronounced among low-income sectors. In general the likelihood of a poor person being computer and Internet literate decreases with age. The difficulty of learning at a later age and the high concentration of illiteracy among older populations cause the digital divide to be particularly large among low-income older adults in developing countries.

The number of computers and users registered to Internet access providers is the principal means for measuring the number of users. For some authors it is necessary to distinguish between active users, for whom the Internet is integrated into daily life, and passive users, for whom Internet use is casual. The diversity of methods of access makes it difficult to count the number of Internet users. There is great disagreement regarding the criteria for evaluating the number of users, even within the United States, between the different organizations and companies specialized in the business. In principle, one assumes that the number of users per access point (computer connected to the Internet) is larger in developing countries than in developed countries and that poor families have more users per computer than rich families (some families have more than one home computer connected to the Internet). In some cases, a single user can be registered with multiple providers, in other cases, as in the tele-centers or cyber-cafes; a single computer provides access to a great numbers of users. When Internet access penetrates the poorest sectors of the population, the number of users per computer tends to increase.

## 4.4 What is the relationship between the Internet and literacy?

**KEYWORDS:**
KNOWLEDGE SKILLS, BOOK LITERACY, EDUCATIONAL INEQUITY

The ability to use the information available on the Internet as a source of knowledge and intellectual and professional development depends on the users' prior skills. This qualification assumes basic literacy and abilities acquired within the school system. **Digital literacy cannot be dissociated from book literacy**. The network multiplies the possibilities for intellectual and professional work but at least until the present moment, it is not a substitute for the basic intellectual qualifications that are acquired at school and its effective potential depends on them. Thus social inequality as expressed in educational inequity is reproduced and increases with use of the Internet. As long as much of the population of the developing world continues to struggle with illiteracy and semi-literacy, universal access to the Internet will be an illusory goal.

Training in the use of the computer and the Internet (called digital literacy or e-literacy) can be offered through formal courses in school or at work, private courses, or courses promoted by non-governmental organizations, or in contexts (schools, work or home) where the Internet is used and people nearby are able to offer assistance when needed. Children, in particular, tend to learn to use computers and the Internet through play almost without direct orientation. However the probability of having the type of access that allows this kind of learning by osmosis, either at home or work, is lower in low-income sectors where the chances of owning a home computer as well as having access to a computer in the work place are extremely low.

The aforementioned factors combine in determining the **uses of** information technology, the most important criteria for evaluating its effective relevance for society. These uses depend on the creative appropriation of the new technology by the different social actors and each user, producing new contents and applications that represent innovative responses to economic, social political, and cultural problems.

The use of the Internet can be analyzed according to its dual dimensions as both an instrument of communication and dissemination of information and an instrument for access to information. It's potential as a communication instrument (email) is greater among high income users since most probably all the members of his/her network have access to the Internet, while this is not the case of the low income users. This is even truer in the case of international contacts, because low income sectors are very unlikely to have an international social network. The only relevant exception relates to poor families with members working abroad, often illegally, for whom the Internet offers cheap communication and contact with their native land and families.

## 4.5 How Internet content is related to social inequality?

**KEYWORDS:**
CONTENT, CONTENT PRODUCTION, CONTENT BIAS, LANGUAGES

The content available on the Internet constitutes a decisive area in the dynamics of the digital divide. The uses that it enables are central factors in the impact of the Internet on social inequality. Even if universal access is assured, the lack of content targeting the poor can limit the effective impact of the Internet on low-income sectors.

Generally, in both developing and developed countries a large part of Internet content is developed for the middle class, the principal market with the potential to indirectly or directly generate revenues for web sites, either through advertising, buying advertised products, or by direct payment for access. The orientation toward middle class users is evident in both form and content as the majority of the sites assume a relatively high user educational level. Even non-commercial home pages tend to be produced by the middle class, since making a web site requires certain knowledge of the Internet or a minimum of financial resources.

The shortage of content specifically created for rural communities is aggravated by the fact that the Internet is basically an urban phenomenon. Users and especially producers of web sites are largely concentrated in large urban centers. In small cities and in low-income neighborhoods of large cities, the production of information concerning local necessities is very limited.

For some time, the main concern associated with the global impact of Internet content production was the predominance of English language sites. In addition to imposing an Anglo-Saxon cultural hegemony, English language sites are socially exclusive as much because of their content (which is not relevant to local conditions), as because they require knowledge of English. In non-English speaking regions, this skill is usually limited to the upper classes.

This concern has been shown to be an exaggeration: as the Internet grows, the percentage of home pages in each language tends to be consistent with the percentage of Internet users (with exception of Asian languages and ethnic minorities). Still international inequalities continue to be important. International Internet traffic indicators show that Latin American and African users consult web sites in advanced countries many times more often than the reverse and, that, while growth in the number of sites in developing countries has accelerated, in comparative terms, most sites still leave much to be desired in terms of quality and the amount of information they offer (this can be easily confirmed by visiting sites of many of the governments of central Africa). The importance of the quantity of information available on English language web sites means that those who lack English language abilities have a limited capacity to make use of the Internet. In the future, instantaneous text translation systems (many of which already are available but do not yet produce high quality translations) will be important instruments in the intra-cultural communication and dissemination of information on a global level.

The existing bibliography indicates a shortage of sites aimed at minority ethnic indigenous groups in developing countries, and has shown that where these sites do exist, they tend to be produced by outside specialists. The shortage of sites dedicated to the needs of poor urban sectors of the population, whose main point of access are the tele-centers, is equally dramatic.

## *4.6 What are the main policies which can diminish the digital divide?*

**KEYWORDS:**
PRICE-SUBSIDY, LOW-COST COMPUTERS, TELE-CENTERS, CIBERCAFÉS, COMMUNITY CENTERS

The policies of infrastructure access expansion, oriented by the privatization policies of the 1990's have been generally successful in expanding the use of the cellular telephone to the low income sectors of society of developing societies. However in relation to Internet access they have collided with the limits of effective demand due. The poorest members of the population do not have the resources to buy a computer and are even less able to pay monthly fees for a telephone line and Internet access provider. The most common mechanisms for addressing this barrier to universal access are: 1) price subsidies for the low-income users, 2) promotion of low-cost or recycled computers, 3) support for the creation of technology centers.

A) Price subsidies – International experience offers examples of reduced access rates in poor neighborhoods, tele-centers with free or subsidized access fees, and subsidized rates that favor low-income users and tele-centers.

B) Promotion of low-cost and recycled computers – The production of "people's computers" has as its main challenge the creation of a product capable of confronting the "double helix" of the computer industry: the need for constant renewal of the hardware due to the increasing demand of information storage and processing of the new software. The main possibility for such a computer to be created would be either through: 1) the development of an alternative computer through joint efforts of a non-profit research center and local firms; or 2) the pro-duction of a cheaper computer by the multi-national industry. In the latter case the problem is that a low-cost computer could take a share away from their existing market. One possible solution would be to target such a device only to the institutional market (schools and public institutions) and tele-centers in low-income areas. In addition to efforts to reduce hardware costs, it would be necessary to find alternative solutions to lowering the costs of the software. The first low-cost computers have failed but new experiences, like the one lead by Nicholas Negroponte are under way. An alternative way to increase access could be the reduction of taxes on basic computers. Recent experience in Brazil has proven to be quite successful in increasing the demand for computers in the lower middle classes.

Still the most important policy to increase access for the poor are collective access points where users can benefit from Internet services using equipment that does not belong to them. Those access points can be tele-centers, with non-profit orientation, or cybercafés, which are profit oriented, and they play a role comparable to that of the public telephone. Tele-centers and cybercafés are the main instrument for advancing universal access in developing countries.

International organizations have developed several typologies of tele-centers. They can be simplified in to the following main types:

**Access providing tele-centers**, provide basic access such as computing, Internet, fax, photocopying, printing and telephone service. In the poorest countries tele-centers often provide only telephone service.

**Single purpose technology centers**; offer a single type of content and services, such as governmental or educational information.

**Training tele-centers**, that include courses in information technology use and user support along with the services mentioned in the first type.

**Multipurpose community tele-centers**, offer several of the following services: local access, information, public services, educational courses, technology courses, community radio, content production, and services for the community.

Despite the importance given to tele-centers, the bibliography details only a small number of documented examples of self-sustainability. Most of the bibliography doesn't go beyond anecdotal material. Still more work is needed to advance tele-centers and cybercafés business models that include technological alternatives, types of software, models of management, systems of payment, services offered, partnerships between the NGOs and the private and public sectors, and forms of integration with the local community.

With the expansion of Internet there is a tendency that the non-profit tele-centers tend to be substituted by privately owned cybercafés. Sometimes based in small spaces or house annexes, administered by the family, or in big cities in more formal spaces, cyber-cafes become the most important way for digital inclusion in most big cities in developing countries. In most of the cases they don't receive direct or indirect support from government and are not considered by regulatory agencies. In general, telecommunications regulating agencies, due to limitations of their mandates or other factors, tend not to regulate access fees that could allow cybercafés in poor regions to reduce access rates.

The models of tele-centers and cybercafes must be adapted to the diverse local contexts, and should be developed directly by the public sector supported by non-governmental organizations and enterprises. The creation of public access points should creatively combine a variety of types of tele-centers.

For example, a model can be imagined where in a particular neighborhood or small village, some collective access points are installed by the private initiative while a public multi-purpose tele-center offers courses, orients the population in Internet use and supports efforts to produce local content with information on the life of the community.

## *4.7 What is e-readiness?*

**KEYWORDS:**
TELECOMMUNICATION PENETRATION, INTERNATIONAL RATING, INTERNATIONAL COMPETITIVENESS

Some international studies seeking to develop indicators for establishing the relative position of countries in terms of information technology development, created the concept of e-readiness. This concept allows evaluating the penetration of communication technologies within countries in comparison with other countries, and is considered to be an important factor in determining international economic competitiveness. The relative e-readiness of a country is not necessarily correlated with the country's internal digital divide. Even so, policies that fight against the digital divide positively affect national capacity in terms of e-readiness.

There are many ways to formulate and define the e-readiness of each country. Some authors identify stages of e-readiness development based on key indicators such as the number of people with access to communication technologies. More complex models consider factors including the institutional contexts, regulatory systems in the area of telecommunications, human resources, systems of innovation, and the uses and impacts on society of new technologies. The simpler formulas suffer from the types of problems that are typical of quantitative comparisons between countries with di-verse economic, political, and socio-cultural realities. This is particularly important in the area where product dissemination and education levels determine the diversity of potential uses for Internet. The more complex formulas also have their shortcomings in that by considering qualitative dimensions, they are more difficult to quantify and they depend on allocating more or less random values to each index.

Despite the different criteria they use, the majority of studies arrive at a typology of levels of e-readiness that confirms strong correlation with the relative international position of the country in terms of per capita income. Still, within each group of countries, there are important differences.

Since Internet use by tourists in developed countries is practically universal, nearly all governments and many hotels in developing countries have web sites aimed at this audience. In many of the poorest developing countries, a large part of publicly and privately run web sites are used mainly for the promotion of tourism. International and national institutions have touted the Internet as an instrument for modernizing small and medium-sized businesses, and as a mechanism for accessing international markets.

We should mention the impact of the new technologies on the **flow of payments between developed and developing countries** and its consequences not only on development but also on the digital divide. While the international telephone payment system favors developing countries, the payment systems for information transfer via the Internet favor developed countries, primarily the United States, the main center of international of Internet traffic.

## *4.8 What are the main public uses of the Internet in developing countries?*

**KEYWORDS:**
DISTANCE EDUCATION, E-GOVERNMENT, E-SCIENCE AND TECHNOLOGY, E-CULTURE, E-HEALTH,

**Distance education** - Distance education precedes the Internet. Correspondence courses followed by radio, television and videocassette, have a long tradition and have served innumerable people, who, either due to time or distance, could not attend a traditional class. In 1969 the Open University, in Great Britain, had a pioneering role in university level education by way of correspondence courses. In the 1980's and 1990's several developing countries created higher education distance education courses, especially for inhabitants of rural areas. Today distance universities in Turkey, China, Indonesia, Thailand, Korea and India have hundreds of thousand of students.

In the majority of distance universities established in developing countries the main means of communication are the post office, radio, television, videotapes and CDs, with the Internet still playing, in general, a supporting role. One of the obvious reasons for the limited use of the Internet is that the majority of poor students lack access. This situation is starting to change and in the past few years, nearly the majority of the principal universities in developing countries have begun to conduct distance education via the Internet.

In primary schools in developing countries, the intensive use of the Internet is mainly concentrated in privates institutions. Since most of the children in developing countries can only access computers in schools, it is fundamental that computers be available to schools even if they are concentrated in collective laboratories. This allows children to become socialized in computer and Internet use, offering a minimum of fa-miliarity with information technology and increasing their future chances in the market place.

Distance education has been defended as a solution for teacher training problems, especially in rural areas. There are several cases in developing countries of the creation of regional training using the Internet and videoconferencing, and "school nets" that offer teachers continuously updated programs and didactic material. The success of these initiatives depends on the availability of Internet access in schools and on the basic training of the teachers in the use of the Internet. Among countries that have made important advancements in the creation of "school nets" are Chile (with the Enlace program supported by a network of universities that already reaches almost all secondary schools and more of the half of primary schools), South Africa, and Thailand .

**E-Science and Technology -** The Internet was originally a tool used by the scientific community and spread rapidly among the most advanced developing countries including the majority of Latin America countries, China, India, the north of Africa and South Africa. It has already spread in the poorest countries, generally with the support of international organizations, foundations, and corporations. For the scientific communities of developing countries, access to the Internet has meant the possibility of easy, fast, and cheap communication with the international scientific community and online access to databases, scientific interchange and participation in specialized international networks and virtual libraries to which they do not have material access. The Internet has an important role in facilitating the dissemination of the intellectual production that otherwise would be effectively lost.

**E-Culture** - In the field of culture one the most important areas of impact of the Internet has been the creation of virtual li-braries, that allow populations of developing countries that are financially unable to construct and to maintain traditional libraries, to access the written assets of humanity. Although access to these texts by computer screen is neither equivalent, nor a substitute to paper based text, virtual libraries allow contact with texts that would otherwise be unavailable to students at schools and universities in the developing world.

The Internet has been a vehicle of cultural globalization as well as an instrument for expression of cultural diversity in the contemporary world as well as making cultural works available around the world. The creation of virtual museums has advanced substantially in some developing countries in particular in Latin America followed by some Asian countries like Korea, China, and Turkey, as well as South Africa.

**E-Health** - The Internet is often presented as the solution for remote regions and/or for regions of developing countries that lack sufficient local medical staffing. Although there is an enormous potential for this use, there are limited possibilities for success in the near future, because the regions with the greatest needs are those with the least access to resources and staff trained for using the instruments of telemedicine, with their continuing high costs. This area continues to be a low-priority for health care systems in developing countries.

The Internet has already been introduced for modernizing administration systems and organization of health care systems in developing countries, leading to more rational resource management and improving the quality of services. Information technology is also used in controlling epidemics, for participation in international monitoring, making the body of medical literature available via virtual libraries,[11] and distributing information between health care professionals, particularly in public health campaigns.

**E-Government** - In developing countries, e-government can be an important instrument for reducing inefficiency and the private appropriation of the state by bureaucracies that transform public services into sources of favors, gratuities and systematic corruption. The ability to access an increasing number of documents and official information via the Internet diminishes the power of bureaucratic and political intermediaries.

However, to the extent that public state services are accessible via the Internet, they tend to create an increasing divide between citizens with and without access. This problem is aggravated with widespread use of public services via the Internet. Since the universalization of Internet access in developing countries will be a long process, it will be necessary to maintain and improve alternative lines of communication between the government and citizens, such as mobile telephone and face to face contact.

## CONCLUSIONS

*What can be done?*

E-social development does not substitute other kinds of social development, nor does the fight against the digital divide substitute the set of measures necessary for facing poverty, social inequality, and one of their most terrible consequences, urban violence. But e-development has become one of the dimensions of social development, as the fight against the digital divide is one of many dimensions of the fight against poverty and inequality. Basic knowledge of ICTs increasingly becomes a precondition for employment. Universalizing basic knowledge of computers and the Internet is fundamental in limiting the negative impact their absence has on the poorest sectors in spite of several limitations of the policies to democratize information.

The struggle for digital inclusion is a struggle against time. New information technologies increase existing social inequalities, therefore policies for digital inclusion are nothing more than a struggle to re-align the possibilities for access to the job market and living conditions. The true value of information depends on the user's ability to interpret it. To be useful, information must be meaningful, must be transformed into knowledge through a process of socialization and practices that build analytical capacities. Therefore confronting the digital divide cannot be separated from confronting the educational divide. Policies to universalize access to the Internet in developing countries will not be successful if they are not associated with other social policies, in particu-lar those relating to education. Obviously, this does not mean that we must wait until we are able to eradicate illiteracy in order to develop digital inclusion policies. The demands of the economy and of job creation require interrelated policies that work with different social sectors and different rhythms in order to universalize public services. At the same time we can not ignore the strong links between different social policies.

The criteria for evaluating efforts in combating the digital divide are how these programs reduce other forms of social inequality and poverty. For example, in the past decade, the United States has practically achieved universal access to the Internet but social inequality has not diminished because of this, in fact it has increased.

The increasing complexity associated with the fight against social inequality creates new challenges for strategic planning of governmental actions and for the development of social policies. Policies aimed at reducing the digital divide are a necessary component of social policy but they are not the answer to all social and economic problems. The same is true in relation to e-education and to the problems brought on by declines in school performance. The introduction of the Internet should be part of the general rethinking of teaching methods and the role of both schools and teachers.

The search for miraculous solutions is constant in developing countries. The Internet is too important to be brought into the cycle of miracle cures that later are abandoned for not meeting unrealistic expectations.

In the first place, developing countries should take the dynamic nature of the digital divide into consideration, which means that countries which are not part of the central nucleus of technology generation develop at least the capacity for **defensive strategic analysis**. This will allow them to follow the trends and experiences developed by information technology leaders in the developing and developed world, thereby reducing the experimentation costs and helping to define the best technological options and most appropriate products for their social realities.

Second, policies must assure the coherence, integration, security and inter-operability of different public services, controlling costs and making government action more coherent. The coordination of the policies cannot be left to short sighted party interests, the institution responsible for defining the strategies for the information society must be part of the central core of government. If the decision making nucleus of the government does not commit to the coordination of ministerial activities associated with the information society, the result will produce waste due to duplicated efforts and unrealistic programs. The fight against the digital divide must be viewed as a long term **state policy**, avoiding the tendency in developing countries for new governments to abandon and devalue the accomplishments of their predecessors.

Third, the coordination of infrastructure development policies is necessary to increase the synergy between different physical networks such as roads, electric lines, telephone and fiber optics. Universal service should be promoted, either by creating incentives so that private companies invest directly in the neediest areas, or government intervening directly to assure services at accessible costs.

Fourth, the urgency to resolve the problem of the digital divide cannot justify hasty pharaonic investments in areas that demand experimental pilot programs, adequate local conditions, user training, systems of evaluation and technical support. This is particularly true of the installation of Internet access in schools, which should be a gradual process. This means that ICTs should not be transformed overnight in a privileged instrument for the educational system through over-invest in exaggerated quantities of computers in each school.

Furthermore, the experience in most developing countries shows that the introduction of computers in public schools is not accompanied by sustained long term investment in equipment maintenance and renewal. The adaptation of professors to this new instrument is a long process that can not be disassociated with the general improvement of professional development. Developing adequate software, adapting pedagogical systems, and developing critical teaching techniques on the use of ICTs will be a necessarily long term learning process in the majority of developing countries. Until that time, the role of ICT labs in schools should be to introduce students to these instruments and their uses and provide them training on basic programs, in order to motivate them to facilitate future insertion in the job market.

Fifth, investments in collective access must be accompanied by preparation of local human resources. For instance, the Internet allows access to excellent educational programs for remote areas, but this possibility does not resolve the problem of shortages of teachers with a minimum education (typical of many remote places). Qualified teachers are necessary for making good use of the available material on the Internet.

Sixth, it is necessary to advance policies that assure the development of tele-centers and cyber-cafes. Specific regulations should ensure that communication services operators and Internet service providers offer public institutions and tele-centers and cybercafés in low-income areas access to infrastructure at reduced cost.

Seventh, the fight against the digital divide requires partnerships between non-governmental organizations, companies, and government, in which the non-governmental organizations and enterprises play an important role as a source of innovation and contracted services. The fight against the digital divide depends, above all, on the capacity of state action to use market impulses and the experiences of non-governmental or-
ganizations and private initiatives to assure that poor sectors of
the population are integrated into and participate in the con-
struction of the global society.

# FURTHER E-READING [12]

The number of sites devoted to the subjects discussed in the book is extremely vast. The following is a partial list of sites where the reader can begin her/his research on information societies and digital divides:

BRIDGES.ORG
http://www.bridges.org/publications

COMMUNITY TECHNOLOGY CENTERS NETWORK
http://www.ctcnet.org/

COMPUTERS FOR AFRICA
http://www.computers4africa.org/

DIGITAL DIVIDE NETWORK
http://www.digitaldividenetwork.org/

DIGITAL DIVIDE.ORG
http://www.digitaldivide.org/

DIGITAL OPPORTUNITY NETWORK
http://www.digitalopportunity.org

GLOBAL CULTURE TRADE AND TECHNOLOGY DIGITAL
DIVIDE PROJECT
http://www.washington.edu/wto/digital/resources.html

INTERNATIONAL TELECOMMUNICATION UNION
http://www.itu.int/home/index.html

INTERNATIONAL DEVELOPMENT RESEARCH
CENTER - IT SITE
http://www.idrc.ca/ev_en.php?ID=26732_201&ID2=DO_TOPIC

THE MORINO INSTITUTE
http://www.morino.org/under_speeches.asp

UNESCO
http://portal.unesco.org/ci/en/ev.php-URL_ID=22981&URL_DO=DO
_TOPIC&URL_SECTION=-465.html

SOCIAL SCIENCE RESEARCH COUNCIL - IT SITE
http://www.ssrc.org/programs/itic/gcsdocs/

THE JOURNAL OF COMMUNITY INFORMATICS
http://ci-journal.net/index.php/ciej

UN DEVELOPMENT PROGRAMME
http://www.undp.org/poverty/library.htm

WOMEN IN GLOBAL SCIENCE AND TECHNOLOGY
http://www.wigsat.org/node/8

WORLD BANK ITC SITE
http://info.worldbank.org/ict/

# Notes

[8] See, in particular, Lawrence Lessig books on the subject:
http://www.lessig.org.

[9] See for instance our work on Rio de Janeiro
shantytowns: Sorj, B, Guedes, L.E., *Internet y pobreza*.
Montevideo: Editora Unesco - Ediciones Trilce, 2005.

[10] The 'new wave' of literature on the digital divides
converges on the need to consider different levels of digital
inclusion. See, inter alia, "Digital Divides: Past, Present and
Future", IT&Society, Vol. 1, Issue 5, Summer 2003 and Mark
Warschauer, "Reconceptualizing the Digital Divide", First
Monday, Vol. 7, no. 7, 2002. Still, there are few quantitative
studies that go beyond the access/non access parameter,
especially for developing countries. See, for the United States,
"The UCLA Internet Report - Surveying the Digital Future,
Year Three", UCLA Center for Communication Policy,
February 2003 (www.ccp.ucla.edu); and on Europe, see,
Eurostat "ICT usage in household and by individuals", 2004,
(http://europa.eu.int/information_society/
activities/statistics/index_en.htm).

[11] See for instance the extremely successful Latin Amer-
ican virtual scientific library of BIREME: www.scielo.org.

[12] I would like to thank Julie Remold for her assistance
in elaborating the Further E-Reading - list.

# LIST OF KEYWORDS

(Some keywords appear duplicated in order to facilitate the search)

Creative Commons Legal Code
Attribution-NonCommercial 3.0 Unported

## License

THE WORK (AS DEFINED BELOW) IS PROVIDED UNDER THE TERMS OF THIS CREATIVE COMMONS PUBLIC LICENSE ("CCPL" OR "LICENSE"). THE WORK IS PROTECTED BY COPYRIGHT AND/OR OTHER APPLICABLE LAW. ANY USE OF THE WORK OTHER THAN AS AUTHORIZED UNDER THIS LICENSE OR COPYRIGHT LAW IS PROHIBITED.

BY EXERCISING ANY RIGHTS TO THE WORK PROVIDED HERE, YOU ACCEPT AND AGREE TO BE BOUND BY THE TERMS OF THIS LICENSE. TO THE EXTENT THIS LICENSE MAY BE CONSIDERED TO BE A CONTRACT, THE LICENSOR GRANTS YOU THE RIGHTS CONTAINED HERE IN CONSIDERATION OF YOUR ACCEPTANCE OF SUCH TERMS AND CONDITIONS.

## 1. Definitions

a) **"Adaptation"** means a work based upon the Work, or upon the Work and other pre-existing works, such as a translation, adaptation, derivative work, arrangement of music or other alterations of a literary or artistic work, or phonogram or performance and includes cinematographic adaptations or any other form in which the Work may be recast, transformed, or adapted including in any form recognizably derived from the original, except that a work that constitutes a Collection will not be considered an Adaptation for the purpose of this License. For the avoidance of doubt, where the Work is a musical work, performance or phonogram, the synchronization of the Work in timed-relation with a moving image ("synching") will be considered an Adaptation for the purpose of this License.

b) **"Collection"** means a collection of literary or artistic works, such as encyclopedias and anthologies, or performances, phonograms or broadcasts, or other works or subject matter other than works listed in Section 1(f) below, which, by reason of the selection and arrangement of their contents, constitute intellectual creations, in which the Work is included in its entirety in unmodified form along with one or more other contributions, each constituting separate and independent works in themselves, which together are assembled into a collective whole. A work that constitutes a Collection will not be considered an Adaptation (as defined above) for the purposes of this License.

c) "**Distribute**" means to make available to the public the original and copies of the Work or Adaptation, as appropriate, through sale or other transfer of ownership.

d) "**Licensor**" means the individual, individuals, entity or entities that offer(s) the Work under the terms of this License.

e) "**Original Author**" means, in the case of a literary or artistic work, the individual, individuals, entity or entities who created the Work or if no individual or entity can be identified, the publisher; and in addition (i) in the case of a performance the actors, singers, musicians, dancers, and other persons who act, sing, deliver, declaim, play in, interpret or otherwise perform literary or artistic works or expressions of folklore; (ii) in the case of a phonogram the producer being the person or legal entity who first fixes the sounds of a performance or other sounds; and, (iii) in the case of broadcasts, the organization that transmits the broadcast.

f) "**Work**" means the literary and/or artistic work offered under the terms of this License including without limitation any production in the literary, scientific and artistic domain, whatever may be the mode or form of its expression including digital form, such as a book, pamphlet and other writing; a lecture, address, sermon or other work of the same nature; a dramatic or dramatico-musical work; a choreographic work or entertainment in dumb show; a musical composition with or without words; a cinematographic work to which are assimilated works expressed by a process analogous to cinematography; a work of drawing, painting, architecture, sculpture, engraving or lithography; a photographic work to which are assimilated works expressed by a process analogous to photography; a work of applied art; an illustration, map, plan, sketch or three-dimensional work relative to geography, topography, architecture or science; a performance; a broadcast; a phonogram; a compilation of data to the extent it is protected as a copyrightable work; or a work performed by a variety or circus performer to the extent it is not otherwise considered a literary or artistic work.

g) "**You**" means an individual or entity exercising rights under this License who has not previously violated the terms of this License with respect to the Work, or who has received express permission from the Licensor to exercise rights under this License despite a previous violation.

h) "**Publicly Perform**" means to perform public recitations of the Work and to communicate to the public those public recitations, by any means or process, including by wire or wireless means or public digital performances; to make available to the public Works in such a way that members of the public may access these Works from a place and at a place individually chosen by them; to perform the Work to the public by any means or process and the communication to the public of the performances of the Work, including by public digital performance; to broadcast and rebroadcast the Work by any means including signs, sounds or images.

i) "**Reproduce**" means to make copies of the Work by any means including without limitation by sound or visual recordings and the right of fixation and reproducing fixations of the Work, including storage of a protected performance or phonogram in digital form or other electronic medium.

**2. Fair Dealing Rights.** Nothing in this License is intended to reduce, limit, or restrict any uses free from copyright or rights arising from limitations or exceptions that are provided for in connection with the copyright protection under copyright law or other applicable laws.

**3. License Grant.** Subject to the terms and conditions of this License, Licensor hereby grants You a worldwide, royalty-free, non-exclusive, perpetual (for the duration of the applicable copyright) license to exercise the rights in the Work as stated below:

a) to Reproduce the Work, to incorporate the Work into one or more Collections, and to Reproduce the Work as incorporated in the Collections;

b) to create and Reproduce Adaptations provided that any such Adaptation, including any translation in any medium, takes reasonable steps to clearly label, demarcate or otherwise identify that changes were made to the original Work. For example, a translation could be marked "The original work was translated from English to Spanish," or a modification could indicate "The original work has been modified.";

c) to Distribute and Publicly Perform the Work including as incorporated in Collections; and,

d) to Distribute and Publicly Perform Adaptations.

The above rights may be exercised in all media and formats whether now known or hereafter devised. The above rights include the right to make such modifications as are technically necessary to exercise the rights in other media and formats. Subject to Section 8(f), all rights not expressly granted by Licensor are hereby reserved, including but not limited to the rights set forth in Section 4(d).

**4. Restrictions.** The license granted in Section 3 above is expressly made subject to and limited by the following restrictions:

a) You may Distribute or Publicly Perform the Work only under the terms of this License. You must include a copy of, or the Uniform Resource Identifier (URI) for, this License with every copy of the Work You Distribute or Publicly Perform. You may not offer or impose any terms on the Work that restrict the terms of this License or the ability of the recipient of the Work to exercise the rights granted to that recipient under the terms of the License. You may not sublicense the Work. You must keep intact all notices that refer to this License and to the disclaimer of warranties with every copy of the Work You Distribute or Publicly Perform. When You Distribute or Publicly Perform the Work, You may not impose any effective technological measures on the Work that restrict the ability of a recipient of the Work from You to exercise the rights granted to that recipient under the terms of the License. This Section 4(a) applies to the Work as incorporated in a Collection, but this does not require the Collection apart from the Work itself to be made subject to the terms of this License. If You create a Collection, upon notice from any Licensor You must, to the extent practicable, remove from the Collection any credit as required by Section 4(c), as requested. If You create an Adaptation, upon notice from any Licensor You must, to the extent practicable, remove from the Adaptation any credit as required by Section 4(c), as requested.

b)   You may not exercise any of the rights granted to You in Section 3 above in any manner that is primarily intended for or directed toward commercial advantage or private monetary compensation. The exchange of the Work for other copyrighted works by means of digital file-sharing or otherwise shall not be considered to be intended for or directed toward commercial advantage or private monetary compensation, provided there is no payment of any monetary compensation in connection with the exchange of copyrighted works.

c)   If You Distribute, or Publicly Perform the Work or any Adaptations or Collections, You must, unless a request has been made pursuant to Section 4(a), keep intact all copyright notices for the Work and provide, reasonable to the medium or means You are utilizing: (i) the name of the Original Author (or pseudonym, if applicable) if supplied, and/or if the Original Author and/or Licensor designate another party or parties (e.g., a sponsor institute, publishing entity, journal) for attribution ("Attribution Parties") in Licensor's copyright notice, terms of service or by other reasonable means, the name of such party or parties; (ii) the title of the Work if supplied; (iii) to the extent reasonably practicable, the URI, if any, that Licensor specifies to be associated with the Work, unless such URI does not refer to the copyright notice or licensing information for the Work; and, (iv) consistent with Section 3(b), in the case of an Adaptation, a credit identifying the use of the Work in the Adaptation (e.g., "French translation of the Work by Original Author," or "Screenplay based on original Work by Original Author"). The credit required by this Section 4(c) may be implemented in any reasonable manner; provided, however, that in the case of a Adaptation or Collection, at a minimum such credit will appear, if a credit for all contributing authors of the Adaptation or Collection appears, then as part of these credits and in a manner at least as prominent as the credits for the other contributing authors. For the avoidance of doubt, You may only use the credit required by this Section for the purpose of attribution in the manner set out above and, by exercising Your rights under this License, You may not implicitly or explicitly assert or imply any connection with, sponsorship or endorsement by the Original Author, Licensor and/or Attribution Parties, as appropriate, of You or Your use of the Work, without the separate, express prior written permission of the Original Author, Licensor and/or Attribution Parties.

d)   For the avoidance of doubt:

   i.   i.Non-waivable Compulsory License Schemes. In those jurisdictions in which the right to collect royalties through any statutory or compulsory licensing scheme cannot be waived, the Licensor reserves the exclusive right to collect such royalties for any exercise by You of the rights granted under this License;

   ii.  e.Waivable Compulsory License Schemes. In those jurisdictions in which the right to collect royalties through any statutory or compulsory licensing scheme can be waived, the Licensor reserves the exclusive right to collect such royalties for any exercise by You of the rights granted under this License if Your exercise of such rights is for a purpose or use which is otherwise than noncommercial as permitted under Section 4(b) and otherwise waives the right to collect royalties through any statutory or compulsory licensing scheme; and,

    iii. f.Voluntary License Schemes. The Licensor reserves the right to collect royalties, whether individually or, in the event that the Licensor is a member of a collecting society that administers voluntary licensing schemes, via that society, from any exercise by You of the rights granted under this License that is for a purpose or use which is otherwise than noncommercial as permitted under Section 4(c).

e) Except as otherwise agreed in writing by the Licensor or as may be otherwise permitted by applicable law, if You Reproduce, Distribute or Publicly Perform the Work either by itself or as part of any Adaptations or Collections, You must not distort, mutilate, modify or take other derogatory action in relation to the Work which would be prejudicial to the Original Author's honor or reputation. Licensor agrees that in those jurisdictions (e.g. Japan), in which any exercise of the right granted in Section 3(b) of this License (the right to make Adaptations) would be deemed to be a distortion, mutilation, modification or other derogatory action prejudicial to the Original Author's honor and reputation, the Licensor will waive or not assert, as appropriate, this Section, to the fullest extent permitted by the applicable national law, to enable You to reasonably exercise Your right under Section 3(b) of this License (right to make Adaptations) but not otherwise.

## 5. Representations, Warranties and Disclaimer

UNLESS OTHERWISE MUTUALLY AGREED TO BY THE PARTIES IN WRITING, LICENSOR OFFERS THE WORK AS-IS AND MAKES NO REPRESENTATIONS OR WARRANTIES OF ANY KIND CONCERNING THE WORK, EXPRESS, IMPLIED, STATUTORY OR OTHERWISE, INCLUDING, WITHOUT LIMITATION, WARRANTIES OF TITLE, MERCHANTIBILITY, FITNESS FOR A PARTICULAR PURPOSE, NONINFRINGEMENT, OR THE ABSENCE OF LATENT OR OTHER DEFECTS, ACCURACY, OR THE PRESENCE OF ABSENCE OF ERRORS, WHETHER OR NOT DISCOVERABLE. SOME JURISDICTIONS DO NOT ALLOW THE EXCLUSION OF IMPLIED WARRANTIES, SO SUCH EXCLUSION MAY NOT APPLY TO YOU.

**6. Limitation on Liability.** EXCEPT TO THE EXTENT REQUIRED BY APPLICABLE LAW, IN NO EVENT WILL LICENSOR BE LIABLE TO YOU ON ANY LEGAL THEORY FOR ANY SPECIAL, INCIDENTAL, CONSEQUENTIAL, PUNITIVE OR EXEMPLARY DAMAGES ARISING OUT OF THIS LICENSE OR THE USE OF THE WORK, EVEN IF LICENSOR HAS BEEN ADVISED OF THE POSSIBILITY OF SUCH DAMAGES.

## 7. Termination

a) This License and the rights granted hereunder will terminate automatically upon any breach by You of the terms of this License. Individuals or entities who have received Adaptations or Collections from You under this License, however, will not have their licenses terminated provided such individuals or entities remain in full compliance with those licenses. Sections 1, 2, 5, 6, 7, and 8 will survive any termination of this License.

b) Subject to the above terms and conditions, the license granted here is perpetual (for the duration of the applicable copyright in the Work). Notwithstanding the above, Licensor reserves the right to release the Work under different license terms or to stop distributing the Work at any time; provided, however that any such election will not serve to withdraw this License (or any other license that has been, or is required to be, granted under the terms of this License), and this License will continue in full force and effect unless terminated as stated above.

## 8. Miscellaneous

a) Each time You Distribute or Publicly Perform the Work or a Collection, the Licensor offers to the recipient a license to the Work on the same terms and conditions as the license granted to You under this License.

b) Each time You Distribute or Publicly Perform an Adaptation, Licensor offers to the recipient a license to the original Work on the same terms and conditions as the license granted to You under this License.

c) If any provision of this License is invalid or unenforceable under applicable law, it shall not affect the validity or enforceability of the remainder of the terms of this License, and without further action by the parties to this agreement, such provision shall be reformed to the minimum extent necessary to make such provision valid and enforceable.

d) No term or provision of this License shall be deemed waived and no breach consented to unless such waiver or consent shall be in writing and signed by the party to be charged with such waiver or consent.

e) This License constitutes the entire agreement between the parties with respect to the Work licensed here. There are no understandings, agreements or representations with respect to the Work not specified here. Licensor shall not be bound by any additional provisions that may appear in any communication from You. This License may not be modified without the mutual written agreement of the Licensor and You.

f) The rights granted under, and the subject matter referenced, in this License were drafted utilizing the terminology of the Berne Convention for the Protection of Literary and Artistic Works (as amended on September 28, 1979), the Rome Convention of 1961, the WIPO Copyright Treaty of 1996, the WIPO Performances and Phonograms Treaty of 1996 and the Universal Copyright Convention (as revised on July 24, 1971). These rights and subject matter take effect in the relevant jurisdiction in which the License terms are sought to be enforced according to the corresponding provisions of the implementation of those treaty provisions in the applicable national law. If the standard suite of rights granted under applicable copyright law includes additional rights not granted under this License, such additional rights are deemed to be included in the License; this License is not intended to restrict the license of any rights under applicable law.

## Creative Commons Notice

Creative Commons is not a party to this License, and makes no warranty whatsoever in connection with the Work. Creative Commons will not be liable to You or any party on any legal theory for any damages whatsoever, including without limitation any general, special, incidental or consequential damages arising in connection to this license. Notwithstanding the foregoing two (2) sentences, if Creative Commons has expressly identified itself as the Licensor hereunder, it shall have all rights and obligations of Licensor.

Except for the limited purpose of indicating to the public that the Work is licensed under the CCPL, Creative Commons does not authorize the use by either party of the trademark "Creative Commons" or any related trademark or logo of Creative Commons without the prior written consent of Creative Commons. Any permitted use will be in compliance with Creative Commons' then-current trademark usage guidelines, as may be published on its website or otherwise made available upon request from time to time. For the avoidance of doubt, this trademark restriction does not form part of the License.

Creative Commons may be contacted at http://creativecommons.org/.

**Publishing studies** series

The book argues ICT are part of the set of goods and services that determine quality of life, social inequality and the chances for economic development. Therefore understanding the digital divide demands a broader discussion of the place of ICT within each society and in the international system. The author argues against the perspectives that either isolates ICT from other basic social goods (in particular education and employment) as well as those that argue that new technologies are luxury of a consumer society. Though the author accepts that new technologies are not a panacea for the problems of inequality, access to them become a condition of full integration of social life. Using examples mainly from Latin America, the work presents some general policy proposals on the fight against the digital divide which take in consideration other dimensions of social inequality and access to public goods.

Bernardo Sorj was born in Montevideo, Uruguay. He is a naturalized Brazilian, living in Brazil since 1976. He studied anthropology and philosophy in Uruguay, and holds a B.A. and an M.A. in History and Sociology from Haifa University, Israel. He received his Ph.D. in Sociology from the University of Manchester in England. Sorj was a professor at the Department of Political Science at the Federal University of Minas Gerais and at the Institute for International Relations, PUC/RJ. The author of 20 books and more than 100 articles, was visiting professor and chair at many European and North American universities, including the Chaire Sérgio Buarque of Hollanda, at the Maison des Sciences de l'Homme, and the Cátedra Simón Bolívar of the Institut des Hautes Études de l'Amérique Latine, in Paris. He is member of the board of several academic journals, advisor to scientific institutions and consultant to international organizations and governments. In 2005 was elected Man Of ideas of the Year. Currently he is professor of Sociology at the Federal University of Rio de Janeiro, Director of the Edelstein Center for Social Research and of the Plataforma Democrática Project, and Coordinator of SciELO Latin American Social Sciences Journals English Edition.

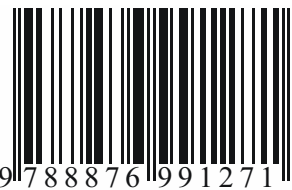

9 788876 991271